\documentclass[useAMS,usenatbib,usegraphicx]{mn2e}
\usepackage{bm}
\usepackage{times}

\title[SMA observation of dwarf galaxies]
{Properties of free--free, dust, and CO emissions
in the starbursts of blue compact dwarf galaxies}
\author[Hirashita]{Hiroyuki Hirashita$^1$\thanks{E-mail:
    hirashita@asiaa.sinica.edu.tw}\\
$^1$Institute of Astronomy and Astrophysics, Academia Sinica,
P.O. Box 23-141, Taipei 10617, Taiwan
}
\date{2012 December 11}

\pagerange{\pageref{firstpage}--\pageref{lastpage}} \pubyear{2012}

\begin{document}
\label{firstpage}
\maketitle

\begin{abstract}
The central star-forming regions in three blue compact
dwarf galaxies (He~2-10, NGC~5253, and II~Zw~40)
were observed in the 340~GHz ($880~\micron$) band
at $\sim 5$ arcsec resolution with the Submillimetre Array
(SMA). Continuum emission associated with the central
star-forming complex was detected in all these galaxies.
The SMA 880 $\micron$ flux is decomposed into free--free
emission and dust emission by using
centimetre-wavelength data in the literature. We find
that free--free emission contributes half or more of
the SMA 880 $\micron$ flux in the central
starbursts in those three galaxies. In spite of the dominance
of free--free emission at 880 $\micron$, the
radio-to-far infrared (FIR) ratios in the central star-forming
regions are not significantly higher than
those of the entire systems,
showing the robustness of radio--FIR relation.
Based on the robustness of the radio--FIR relation,
we argue that the free--free fraction in the 880 $\micron$
emission is regulated by the dust temperature.
We also analyze the CO ($J=$ 3--2) emission data.
We find that CO is a good tracer of the total gas mass
in solar-metallicity object He 2-10.
Low-metallicity objects, NGC~5253 and II~Zw~40, have
apparently high star formation efficiencies; however, this
may be an artifact of significant dissociation of CO
in the low-metallicity environments.
We also point out a potential underestimate of dust mass,
since the dust traced by emission is biased to
the most luminous high-temperature regions, particularly
when a system hosts a compact star-forming region
where the dust temperature is high.
\end{abstract}


\section{Introduction}

The early stage of galaxy evolution can be
characterized by the following two properties:
a poor metal abundance and a rich gas content.
Although galaxies at the early evolutionary stages
exist at high redshift, nearby blue compact dwarf
galaxies (BCDs) are a unique category of galaxies that
have those two properties in the nearby Universe
\citep*{sargent70,vanzee98,kunth00}.
Some BCDs are also experiencing the most active class
of star formation with the formation of
super star clusters (SSCs)
\citep{turner98,kobulnicky99}.
This kind of intense star formation may also provide
a relevant laboratory to understand the conditions in which
high-redshift galaxies experience their first starburst episode.

Such intense star formation as seen in BCDs is occurring
in dense and compact regions and thus can
only be traced with optically thin star formation
indicators, such as far-infrared (FIR) dust luminosity
\citep[e.g.][]{kennicutt98,inoue00} and radio luminosity
(thermal free--free emission from H \textsc{ii}
regions plus non-thermal synchrotron emission
from accelerated electrons)
\citep{condon92}.
These two luminosities are strongly correlated in
nearby star-forming galaxies
\citep[e.g.][]{dejong85,helou85}.

In our previous paper,
\citet[][hereafter H11]{hirashita11}, we observed
a well studied BCD, II~Zw~40, {with}
the Submillimetre Array (SMA; \citealt{ho04}).
In this paper, we add two BCDs, He 2-10 and NGC~5253,
both of which also host high star formation activities
likely to be associated with the formation of SSCs
in the centre \citep{gorjian96,johnson00}.
The radio emission at wavelengths $\ga$ a few cm
is optically thick {for free--free absorption},
supporting the existence of young
compact regions before the expansion of H \textsc{ii}
regions
\citep*{turner98,kobulnicky99,johnson03}.\footnote{{In
this paper, we focus on
the wavelength range ($\leq 2$ cm), where the radiation is optically
thin for free--free absorption as far as our sample is concerned
(Sections \ref{subsec:he2-10}--\ref{subsec:iizw40}).}}
All these BCDs are also classified as Wolf-Rayet galaxies:
the Wolf-Rayet feature
indicates that the
typical age of the current starburst is a few Myr
\citep{vacca92,lopez10}.

Some of the conclusions derived from II~Zw~40 by H11
can be generalized by increasing the sample.
H11 argues that free--free dominated submillimetre
emission can be a characteristics of young active
star formation. H11 has clarified that, if we focus on the
central star-forming region in II~Zw~40,
free--free emission dominates even at
880~$\micron$. Considering that
global\footnote{In this paper, we call the
luminosity of the entire galaxy `global
luminosity'.} submm luminosity is
usually dominated by dust {\citep{galliano05}},
the dominance of
free--free emission at 880 $\micron$ is a special
characteristics, which should be examined with
a larger sample.

By increasing the sample, we can also {explore}
the metallicity dependence of various properties.
The metallicity of the interstellar medium is a
fundamental quantity characterizing the galaxy
evolution because it reflects the enrichment in
heavy elements by stellar generations.
Some metallicity effects are expected in the formation
of dense star-forming regions: (i)
Dust-to-gas ratio in low metallicity objects
is generally low
{\citep{schmidt93,lisenfeld98,hirashita_etal02}}, which implies
that the star-forming regions are less embedded by
dust in low-metallicity objects than in high-metallicity
objects; and (ii) in a
low-metallicity (i.e.\ dust-poor) environment,
star formation is expected to be less efficient
because of less shielding of ultraviolet (UV)
heating photons by dust and molecular hydrogen
\citep{hirashita02,yamasawa11,gnedin11}.
Thus, we will address the
metallicity dependence of dust abundance and
star formation efficiency by investigating gas and
dust emission in BCDs.

In observing intense starbursts associated with
the formation of SSCs, a high spatial resolution is
crucial to spot the young star-forming component.
While high-resolution data of the central star-forming
regions in BCDs are available through various radio
interferometric observations with arcsecond resolutions
(1 arcsec corresponds to 51 pc for He 2-10 and
II Zw 40, and 18 pc for
NGC 5253), information of FIR--submm
dust emission on
such a small scale was lacking for BCDs.
We thus performed SMA observations of
a few BCDs to resolve their submm emission.
Shorter wavelengths such as mid-infrared can achieve
almost arcsecond resolutions by single-dish telescope
facilities such as the \textit{Spitzer Space Telescope}
Infrared Array Camera \citep{fazio04} and \textit{AKARI}
Infrared Camera \citep{onaka07},
but the mid-infrared emission is dominated by
stochastically heated
very small grains \citep{draine85}, which are not
representative of the total dust amount
(although it is empirically known that the mid-infrared
luminosity is well correlated
with the total dust luminosity
in BCDs (\citealt{wu08}; see also \citealt{takeuchi05})).

This paper is organized as follows. We explain the
observations and the data analysis in
Section~\ref{sec:obs}. In Section~\ref{sec:model},
we derive basic quantities related to star formation
and submm emission, and discuss radio--FIR relation.
We also analyze CO(3--2) emission and derive
the molecular gas mass, which is used to estimate
the dust-to-gas ratio and the star formation efficiency.
After discussing our observational results in
Section~\ref{sec:discussion},
we conclude in Section \ref{sec:conclusion}.

\section{Observations and data analysis}\label{sec:obs}

We selected nearby BCDs which host young compact
starbursts. Searching for objects feasible for the sky
coverage and sensitivity of SMA,
we chose He~2-10 and NGC 5253. They have similar
properties to our previous {object}, II~Zw~40, in H11:
(i) The ages of the central starbursts are young
($\la$ a few Myr) as indicated by the {Wolf-Rayet}
feature \citep{vacca92,lopez10}
and the stellar spectral synthesis models
\citep{chandar05}; and (ii) they host optically thick
compact free--free emission
{at wavelengths $\ga$ a few cm}, indicating
an intense starburst in a dense compact region
\citep{turner98,kobulnicky99,johnson03}.
Since He 2-10 has a higher metallicity
than the other two galaxies, we {may be able to
obtain a hint for
metallicity effects} (Section \ref{sec:model};
{see Table \ref{tab:data} for the observed metallicity
values}).
The distances ($D$) and velocities
(relative to the local standard of rest; $V_\mathrm{LSR}$)
adopted in this paper are listed in Table \ref{tab:data}.

The SMA observations of He 2-10 and NGC 5253 were
carried out in the 340 GHz (880 $\micron$) band on
2012 January 11 in the
subcompact configuration. Seven antennas were used with
projected antenna separations between 9.5 and 45 m. The
receivers have two sidebands, the lower and upper
sidebands, which covered the frequency ranges from 330.8
to 334.8 GHz,
and from 342.8 to 346.9 GHz, respectively. The visibility
data were calibrated with the {\footnotesize MIR} package. As a
flux calibrator we used Callisto
(with an adopted flux of 14.5 Jy) for He 2-10, and Titan
(with an adopted flux of 2.36 Jy) for NGC 5253. We used
quasars J0730$-$116 and J1337$-$129 as amplitude and
phase calibrators for
He~2-10 and NGC 5253, respectively.
We adopted quasar 3C279 as a band pass
calibrator for both objects.
In generating the continuum data, we excluded the chunk
containing the CO(3--2) emission at rest 345.796~GHz.
The calibrated visibility data were imaged
and CLEANed with the {\footnotesize MIRIAD} package. The synthesized
beam has a full width at half-maximum (FWHM) of
{5.7 arcsec $\times$ 4.4 arcsec}
($290~\mathrm{pc}\times 220~\mathrm{pc}$)
with a major axis position angle of $41\degr$ for
He 2-10 and a FWHM of
{6.7 arcsec $\times$ 4.1 arcsec}
($120~\mathrm{pc}\times 74~\mathrm{pc}$)
with a major axis position angle of $33\degr$ for
NGC 5253.
The largest angular scale sampled by this observation
is 19 arcsec.
We also use the SMA subcompact data of II Zw 40 in H11
(see H11 for details).

\begin{table*}
\centering
\begin{minipage}{175mm}
\caption{Data.}
\label{tab:data}
\begin{tabular}{lccccccccc}
\hline
Object & R.A. & Dec. & $D\,^\mathrm{a}$ &
$V_\mathrm{LSR}\,^\mathrm{b}$ &
peak flux$^\mathrm{c}$
& total flux$^\mathrm{c}$ & free--free$^\mathrm{d}$ & fraction$^\mathrm{e}$
& $12+\log (\mathrm{O/H})^\mathrm{f}$\\
 & (J2000) & (J2000) & (Mpc) & km s$^{-1}$ & (mJy/beam) & (mJy) &
(mJy) & & \\
\hline
He 2-10  & $08^\mathrm{h}36^\mathrm{m}15\fs 1$ &
$-26\degr 24\arcmin 33\farcs 8$ & 10.5 & 850 & $12.2\pm 1.6$ &
$26.2\pm 2.3$ & 12.8--15.4 & 0.45--0.64 & 8.93\\
NGC 5253 & $13^\mathrm{h}39^\mathrm{m}55\fs 9$ &
$-31\degr 38\arcmin 24\farcs 4$ & 3.7 & 401 & $37.9\pm 2.6$ &
$67.0\pm 3.3$ & $48\pm 4$ & 0.63--0.82 & 8.14\\
II Zw 40 & $05^\mathrm{h}55^\mathrm{m}42\fs 6$ &
$+03\degr 23\arcmin 31\farcs 8$ & 10.5 & 767 & $8.43\pm 1.15$ &
$13.6\pm 2.0$ & {10--13.5} & {0.64--1} & 8.13\\
\hline
\end{tabular}

\medskip

$^\mathrm{a}$ Distance. References: \citet{tully88} for
He 2-10; \citet{caldwell89} for NGC 5253;
H11 for II Zw 40.\\
$^\mathrm{b}$ Velocity relative to the local standard
of rest. References: \citet{vanzi09} for
He 2-10 and NGC 5253; \citet{sage92} for II Zw 40.\\
$^\mathrm{c}$
{Peak and total continuum fluxes of our
SMA 880 $\micron$ observations.
The errors are estimated from the 1 $\sigma$
background fluctuation.}\\
$^\mathrm{d}$ The contribution from free--free emission
at 880 $\micron$.\\
$^\mathrm{e}$ The fraction of free--free contribution.\\
{$^\mathrm{f}$ Oxygen abundance in the literature:
\citet{kobulnicky_etal99} for He 2-10 and NGC 5253;
\citet{thuan05} for II Zw 40.}
\end{minipage}
\end{table*}

\begin{figure}
\includegraphics[width=0.47\textwidth]{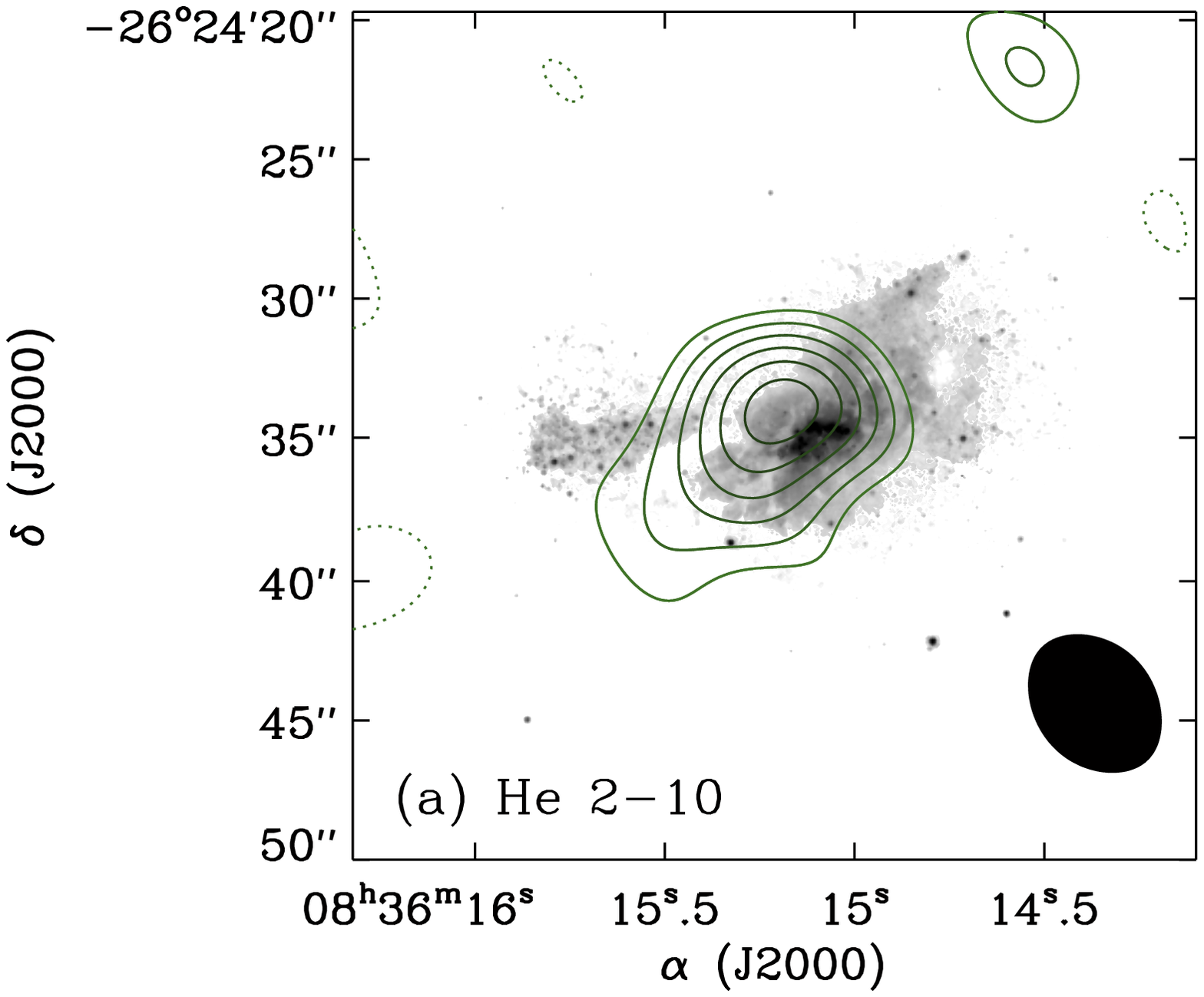}
\includegraphics[width=0.47\textwidth]{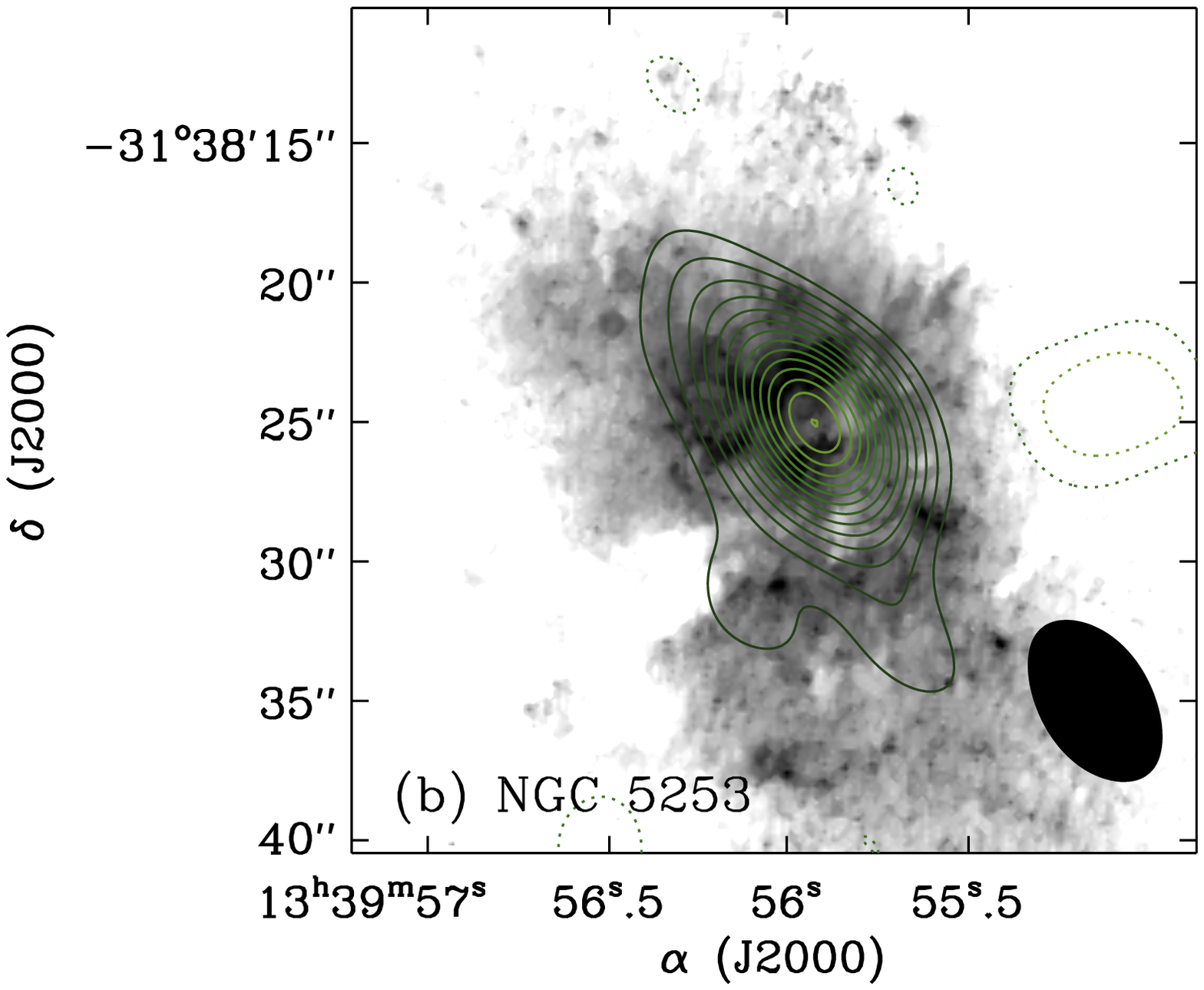}
\includegraphics[width=0.47\textwidth]{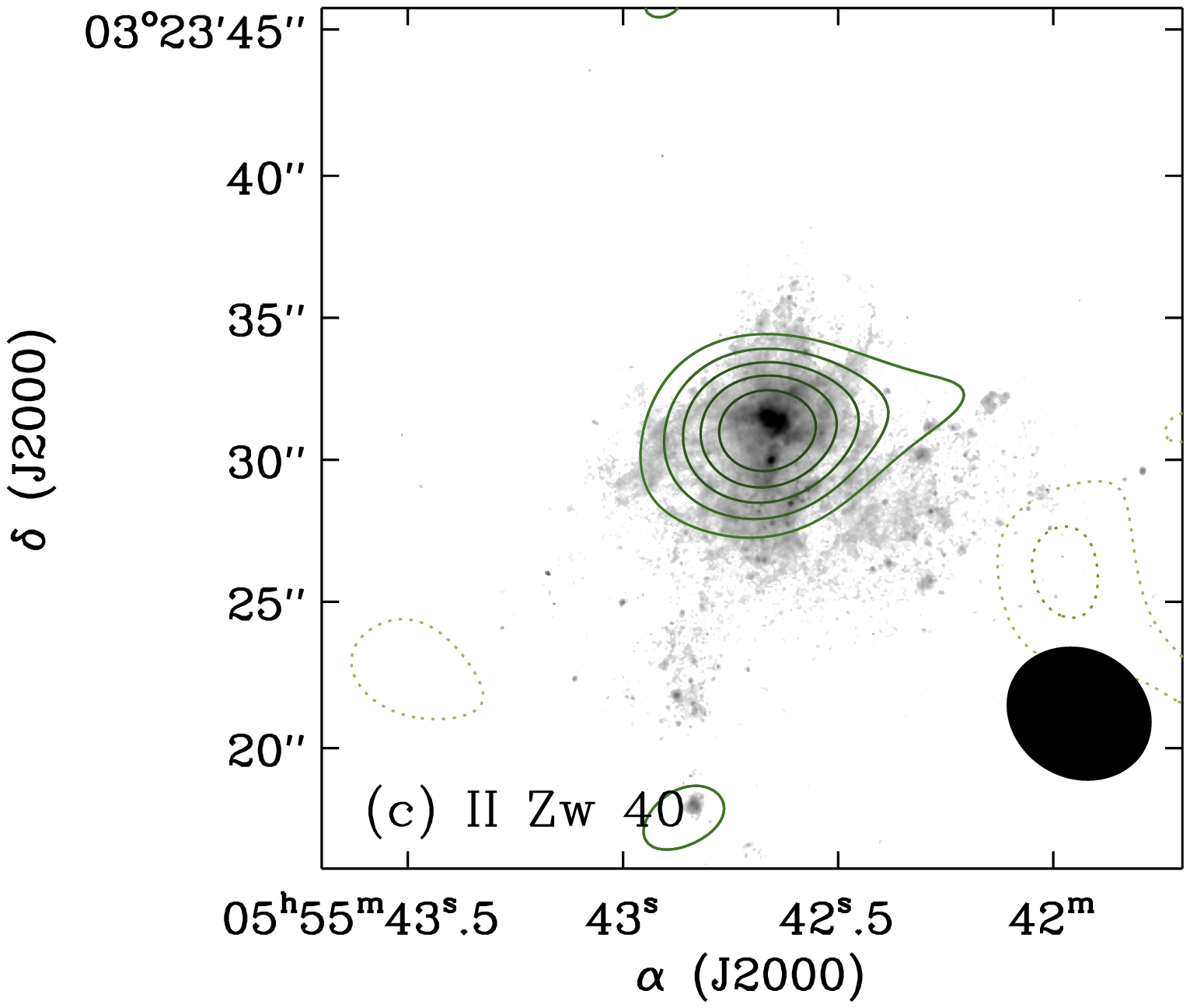}
\caption{{Contour: SMA 880 $\micron$ continuum brightness}
of (a) He 2-10, (b) NGC 5253, and (c) II Zw 40.
Solid contours are 2 $\sigma$, 3 $\sigma$, ...
(1 $\sigma =1.15$~mJy beam$^{-1}$), while dotted
contours are $-2$ and $-3$ $\sigma$. The beam is
shown in the lower right corner.
{Grey scale: \textit{HST} ACS optical images.
at \textit{F550M}, \textit{F555W}, and
\textit{F550W} bands for He 2-10, NGC 5253, and II Zw 40,
respectively.} \label{fig:image}}
\end{figure}

Figure \ref{fig:image} shows the obtained
continuum brightness distribution.
The central active star-forming regions are
detected in all the galaxies. The peak fluxes with
1 $\sigma$ errors are listed in Table \ref{tab:data}.
All the sources are more extended than the beam
{($\sim 5$ arcsec)},
and the total fluxes are also listed in
Table~\ref{tab:data}. {The high brightness
region ($>2\sigma$) is concentrated, considering
that the interferometry is sensitive to an
extension of 19 arcsec. Hereafter, we use the
term `central star forming-region' to indicate the
region where the
SMA brightness is $>2\sigma$. If we compare the
SCUBA images of II Zw 40 and He 2-10 at 450 $\micron$
(beam size $\sim$ 8.5 arcsec)
in \citet{galliano05}, the size of the centrally
concentrated bright regions are consistent, but we
miss the diffuse component (shown by the lowest
contours extending $\ga 1$ arcmin in their
figure 1). The diffuse component is also clear
in their 850 $\micron$ image with a larger beam
($\sim 15$ arcsec).}
We mainly analyze and discuss
the {SMA} continuum data, although we additionally use
the CO(3--2) data in Section \ref{subsec:co}.

In Fig.~\ref{fig:image}, the SMA 880 $\micron$ brightness
distribution is overlaid with
the \textit{Hubble Space Telescope} (\textit{HST})
Advanced Camera for Surveys (ACS) data in the optical
(at \textit{F550M}, \textit{F550W}, and
\textit{F550W} bands for He 2-10, NGC 5253, and II Zw 40,
respectively)\footnote{http://hla.stsci.edu/}.
The positional accuracy of our
SMA image is $\la 0.1$ arcsec, while that of
the \textit{HST} is $\sim 0.5$ arcsec, limited by
the positional uncertainties of the guide stars
\citep{lasker90}.
The centre of submillimetre emission
coincides with that of optical for II Zw 40, while
there is a significant shift in He 2-10. As shown by
\citet{kobulnicky99}, the 3.6 cm radio image also shows
a poor correlation with the \textit{HST} optical image
for He 2-10.
The poor positional coincidence can be attributed to the
extinction in the optical image. The optical image of
NGC 5253 has a complicated morphology, and the
peak of the SMA image does not match the brightest
region in the optical. The different morphologies
between optical and submm (or radio) brightnesses,
especially for He 2-10 and NGC 5253, indicate that
the optical light may be tracing completely different
regions from those observed in submm and radio.

Submm continuum emission in galaxies is usually
dominated by dust thermal radiation and
contaminated by free--free emission
\citep[e.g.][]{galliano05}. Both these emission
processes trace star-forming regions
\citep[e.g.][]{condon92}.
In order to separate the free--free contribution at
880 $\micron$,
we compare our submm fluxes with radio
interferometric fluxes in the literature as explained
in the following subsections.
{We neglect the non-thermal component
throughout this paper since the radio spectrum
at wavelengths $\la$ a few cm is consistent with
free--free emission in the central
star-forming regions (see each subsection below).}

\subsection{He 2-10}\label{subsec:he2-10}

\citet{kobulnicky99} observed He 2-10 by the
Very Large Array (VLA). Their 3.6 cm continuum
image (with a beam size of
0.70 arcsec $\times$ 0.57 arcsec FWHM) fairly
traces our SMA brightness, although individual knots
detected in the VLA image are not resolved by our
SMA observation. They also have high-resolution
images at 2 and 6 cm, where the measured fluxes are
$17.5\pm 0.7$ mJy and $14.5\pm 0.3$ mJy, respectively.
The 3.6 cm flux whose $(u,\, v)$ coverage is matched to
these maps is $15.6\pm 0.4$ mJy. These three fluxes are
consistent with the flat spectrum expected for
free--free emission
\citep[see also][]{kobulnicky99}. Since the spectrum
at 2--6 cm is flat, a large free--free absorption at
this wavelength range can be rejected.
Therefore, we use the interferometric 2 cm flux
($17.5\pm 0.7$ mJy) to extrapolate
the contribution
from free--free emission at 880~$\micron$.
The flux estimated in this way may be a lower limit,
since these matched
data are sensitive to the compact knots and
the structures surrounding them, but possibly miss
the diffuse south-east extension in our SMA image
(Fig.\ \ref{fig:image}). This extension is just along
the possible tidal tail pointed out by a CO
observation in \citet{kobulnicky95}.
Thus, we also use the total VLA flux
(21.1~mJy at 2~cm) as an upper limit. If we convert
these fluxes
to 880~$\micron$ fluxes by assuming $\nu^{-0.1}$
dependence \citep{osterbrock89},
we obtain 12.8 mJy and 15.4 mJy
for the lower and upper limits of the free--free
contribution, respectively. These values indicate
that 45--64 per cent of the SMA 880~$\micron$ flux is
free--free emission.

\subsection{NGC 5253}

\citet*{meier02} show that the
radio continuum spectral slope of NGC 5253 is
consistent with free--free emission.
The slight extension to the west of the peak
is common between their 3.1 mm image and
our SMA 880 $\micron$ image. We adopt
the total flux in the inner 20 arcsec region
($54\pm 5$ mJy at 3.1 mm;
\citealt{meier02}),
to which SMA can be sensitive.
The flat spectrum at $\ga 2$ cm show that
free--free emission is optically thin at 3.1 mm.
We estimate the
880 $\micron$ free--free flux by assuming a frequency
dependence of $\nu^{-0.1}$ for the
free--free spectrum. Then we obtain $48\pm 4$ mJy
for the contribution from free--free emission
at 880 $\micron$. Thus, 63--82 per cent of the
SMA 880 $\micron$ flux is free--free emission.

\subsection{II Zw 40}\label{subsec:iizw40}

This galaxy has already been analyzed and reported
in H11. The centimetre emission from the central
part can be fitted by the free--free flat spectrum.
Since the VLA 2 cm observation is only sensitive to
the structures smaller than 4 arcsec, we also adopt
the single dish flux at 2 cm (18.5 mJy;
\citealt{beck02}) as an upper limit.
H11 also justifies the wavelength dependence
of $\nu^{-0.1}$ at $\la 2$ cm for free--free
emission by spectral fitting.
If we adopt 14--18.5 mJy
for 2 cm flux, we obtain 10--13.5 mJy for 
the contribution from free--free emission
at 880 $\micron$.
Thus, 64--100 per cent of the
flux detected at 880 $\micron$ by SMA is free--free
emission from the central star-forming region.


\section{Models}\label{sec:model}

In order to interpret the 880 $\micron$ emission
in our sample, we need to model two major
emission mechanisms (thermal free--free
radiation and dust emission)
from the star-forming regions. In particular,
the total luminosity of dust emission (called FIR
luminosity) is used to investigate the radio--FIR
luminosity relation later. 
Physical quantities governing
free--free and dust emissions are introduced.
{One of the basic quantities is}
the stellar mass ($M_*$) formed
at the current episode of star formation, since
the ionizing photon luminosity which determines the
free--free emission and the UV luminosity which
contributes to the heating of dust are proportional
to $M_*$.
To evaluate $M_*$, the free--free fluxes
extrapolated from radio observations in
Sections~\ref{subsec:he2-10}--\ref{subsec:iizw40}
are used (Section \ref{subsec:ff}). The obtained
$M_*$ is later used to estimate the UV luminosity
in Section \ref{subsec:dust}.
In addition, we need to determine
the mass and temperature of the dust to estimate
the FIR luminosity, which is used to examine the
radio--FIR relation. Through the modeling,
we can obtain the stellar mass formed in the
current starburst episode, the dust mass
(the dust optical depth), the
dust temperature, all of which are basic quantities
to understand the strength of starburst and
the extent of dust enrichment.
We use the
theoretical models that we applied to II Zw 40
in our previous paper (H11).
Some detailed assumptions that do not affect our
results are simplified.
Below we briefly summarize the models used in
this paper. The same parameter values as those
in H11 are adopted unless otherwise
stated.

\subsection{Basic setups for the star formation}
\label{subsec:sfr}

In H11, we modeled the star formation rate (SFR)
through the free-fall time-scale under
a given gas density, while in this paper, we treat
the SFR as a free parameter. This is because
the SFR is more directly connected to
the observed luminosity than the free-fall time and
the gas density. We assume that the SFR is constant
as a function of
time. As shown below, since we only consider young
($\la 3$ Myr)
star-forming regions, the dependence of the
luminosity on star formation history is weak
in the sense that the total stellar luminosity is
simply determined by the total stellar mass formed
in the current star formation episode.
We assume a
Salpeter initial mass function (IMF) with a stellar
mass range of 0.1--100~M$_{\sun}$.

\subsection{Thermal free--free emission}\label{subsec:ff}

Free--free emission contains the information of
the total stellar mass formed in the current starburst
episode as modeled below. We use the free--free
emission as estimated in
Sections \ref{subsec:he2-10}--\ref{subsec:iizw40}.
The total stellar mass
is necessary to estimate the UV luminosity based on
which the FIR luminosity is modeled in Section
\ref{subsec:dust}.
Thus, we first relate the free--free emission to
the total stellar mass.
{We assume that free--free emission
is optically thin since
we only consider high frequencies such as $\ga 15$ GHz
(see Sections \ref{subsec:he2-10}--\ref{subsec:iizw40}
for further justification).}

The thermal free--free luminosity is proportional
to the number of ionizing photons emitted per unit
time, $N_\mathrm{ion}$ \citep{condon92}. We relate
the SFR with $N_\mathrm{ion}$ following
Section 4.1 of H11 (originally, equation 1 of
\citealt{hirashita06}).
In this paper, we adopt the ionizing photon
luminosity as a function of stellar mass by taking
the solar metallicity case in \citet{schaerer02}.
The dependence of $N_\mathrm{ion}$
on the stellar metallicity is within a factor of 2 in
the metallicity range concerned in this paper.

\begin{figure}
\includegraphics[width=0.48\textwidth]{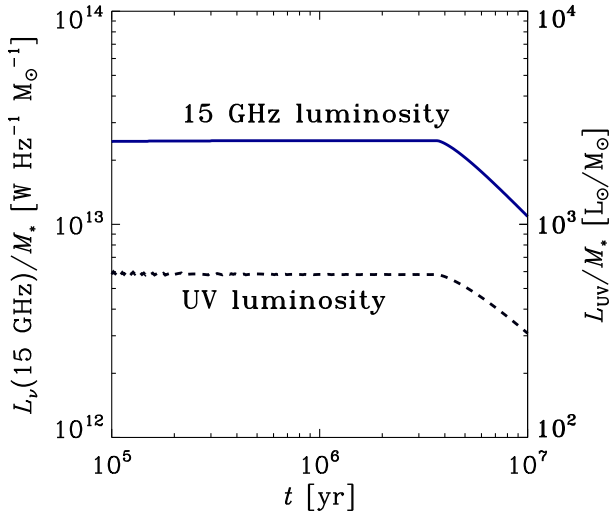}
\caption{Time evolution of the free--free luminosity
at 15 GHz (2 cm)
(solid line; left axis for the scale) and the UV
luminosity (dashed line; right axis for the scale).
Both luminosities are normalized to the stellar mass
formed. For age $\la 3$ Myr, the decline of
luminosities by stellar death
is negligible so that both luminosities normalized
to the stellar mass are constant.
\label{fig:lum}}
\end{figure}

Because we mainly consider young ($\la 3$ Myr)
galaxies, the decline of luminosities by stellar death
is small. Thus, the luminosity basically reflects
the total stellar mass formed up to the current
age of the system.
In Fig.\ \ref{fig:lum}, we show the evolution of the
free--free luminosity ($L_\nu$) at 15 GHz normalized to
the stellar
mass formed (integration of the SFR over the time),
which is denoted as $M_*$. Although we only show
the results at 15 GHz, $L_\nu$ at other frequencies can
be estimated by assuming a frequency dependence
of $\propto\nu^{-0.1}$ as long as
the emission is dominated by free--free and is
optically thin
{(Sections \ref{subsec:he2-10}--\ref{subsec:iizw40})}.
We assume a constant SFR, but the
following results are not sensitive to
the time variation of the SFR under a given $M_*$
as long as
we consider young ($\la 3$ Myr) ages. We obtain the
following formula:
\begin{eqnarray}
\left(
\frac{M_*}{\mathrm{M}_{\sun}}\right) =
\left[
\frac{L_\nu (15~\mathrm{GHz})}
{2.45\times 10^{13}~\mathrm{W~Hz^{-1}}}\right] ,\label{eq:ff}
\end{eqnarray}
where $L_\nu (15~\mathrm{GHz})=4\pi D^2f_\nu
(15~\mathrm{GHz})$
[$f_\nu (15~\mathrm{GHz})$ is the flux at 15 GHz (2 cm) 
adopted in Sections \ref{subsec:he2-10}--\ref{subsec:iizw40}
and listed in Table~\ref{tab:quantities}].
This equation is not valid if there is a significant
contribution from non-thermal synchrotron emission or
if the age is much older than 3 Myr. In
our SMA sample, the contribution from non-thermal
emission should be small since the observational
spectral indexes at centimetre wavelengths are
consistent with thermal free--free (Section \ref{sec:obs}).

In summary, using $L_\nu (15~\mathrm{GHz})$ estimated
from $f_\nu (15~\mathrm{GHz})$ {in
Sections \ref{subsec:he2-10}--\ref{subsec:iizw40}}
(Table \ref{tab:quantities}), we obtain
the stellar mass formed in the current starburst
episode ($M_*$) by equation (\ref{eq:ff}).
The obtained $M_*$ is used {in Section \ref{subsec:dust}}
to estimate the UV luminosity.

\subsection{Dust emission}\label{subsec:dust}

To give a physical interpretation to the dust emission
component in the observed 880 $\micron$ flux,
we need to determine the mass and temperature of
the dust. The dust mass (the dust optical depth) and
the dust temperature are basic quantities to
understand the strength of starburst and the extent
of dust enrichment.  The total FIR luminosity can also
be derived and used to examine the radio--FIR relation
(Section \ref{subsec:radio_FIR}). We do not directly use
optical and UV observations, since they may only trace
regions with
less extinctions and may not be relevant to what
we are modeling (Section \ref{sec:obs}).
We use the same simple model as H11
for the dust emission associated with
the central star-forming region in each individual
galaxy. Here we summarize the model.
At submm wavelengths, the `large' grains, which
achieve radiative
equilibrium with the ambient stellar radiation field,
are the dominant component in the luminosity
\citep[e.g.][]{galliano05}.
{For simplicity, we assume that
the dust is distributed in
a thin shell
at a distance $R_\mathrm{dust}$ from the centre
and that the young stars are located at the centre of the
shell (as also assumed in \citealt{galliano05}).
The following results are valid as long as
the starbursts are more concentrated than the
dust distribution. In our sample, the stars associated
with the current starburst are more compact than
the dust distribution \citep{gorjian96,johnson00,vanzi08}.
In reality, we should also note that it is difficult to
constrain the spatial distribution
of stars because of large dust extinctions.
}

{Under the above assumptions,}
the dust optical depth, $\tau_\mathrm{dust}$, for the
radiation from stars is estimated as
\begin{eqnarray}
\tau_\mathrm{dust}=
\frac{3M_\mathrm{dust}}{16\pi R_\mathrm{dust}^2as},
\label{eq:taudust}
\end{eqnarray}
where $M_\mathrm{dust}$ is the total dust mass
in the shell, $a=0.1~\micron$ is the typical
grain radius (we adopt the geometrical cross section
for the absorption cross section), and
$s=3$ g cm$^{-3}$ is the
grain material density \citep[e.g.][]{draine84}.

We assume that the UV luminosity (denoted as
$\mathcal{L}_\mathrm{OB}$) is equal to the
bolometric luminosity of the OB stars (stars
heavier than 3~M$_{\sun}$); that is, the total
UV luminosity is calculated by summing all the
contribution from OB stars under the star
formation history and the initial mass function
given in Section \ref{subsec:sfr}.
In Fig.~\ref{fig:lum}, we show the evolution of
the UV luminosity calculated by the model
(see H11 for details). Since \citet{schaerer02}
does not
provide the stellar luminosities for solar metallicity,
we adopt the zero-metallicity case. According to
\citet{raiter10}, the variation of
UV luminosity by metallicity under a fixed SFR is
within 0.2 dex if the Salpeter IMF is
adopted. Following the result shown in
Fig.\ \ref{fig:lum}, we obtain
\begin{eqnarray}
\mathcal{L}_\mathrm{OB}=5.8\times 10^2\left(
\frac{M_*}{\mathrm{M}_{\sun}}
\right)~\mathrm{L}_{\sun},\label{eq:LOB}
\end{eqnarray} 
where $M_*$ is already given by the observed
radio flux through equation (\ref{eq:ff}).
We assume that the OB stars are located at the
centre,
so that we can estimate the total FIR
luminosity, $L_\mathrm{FIR}$, by
\begin{eqnarray}
L_\mathrm{FIR}=\left( 1-e^{-\tau_\mathrm{dust}}\right)
\mathcal{L}_\mathrm{OB},
\end{eqnarray}
where $\tau_\mathrm{dust}$ is given by
equation (\ref{eq:taudust}).
The dust temperature, $T_\mathrm{dust}$,
is related to the dust mass, $M_\mathrm{dust}$,
as
\begin{eqnarray}
L_\mathrm{FIR}=1.09\times
10^{-5}M_\mathrm{dust}T_\mathrm{dust}^6,\label{eq:dust_temp}
\end{eqnarray}
The monochromatic dust emission flux is expressed as
\begin{eqnarray}
f_\mathrm{dust}(\nu )=\kappa_\nu M_\mathrm{dust}B_\nu
(T_\mathrm{dust})/D^2,\label{eq:dustsed}
\end{eqnarray}
where $B_\nu$ is the Planck function, $\kappa_\nu$ is
the mass absorption coefficient
of the dust, and $T_\mathrm{dust}$ is the dust
temperature. We assume that
$\kappa_\nu =0.7(\nu /340~\mathrm{GHz})^2$
cm$^2$ g$^{-1}$ \citep{james02}.

The flux from dust $f_\mathrm{dust}(\nu )$ at
880~$\micron$ ($\nu =340$ GHz) used in
equation (\ref{eq:dustsed}) is obtained by
subtracting the free--free flux from the total SMA flux
as described in Section \ref{sec:obs}.
The values of $f_\mathrm{dust}(\nu )$ at
880~$\micron$ are listed in Table \ref{tab:quantities}.
Considering the errors, the upper/lower value of
the free--free contribution and the lower/upper
value of the SMA 880 $\micron$ flux are used to
obtain the lower/upper values of $f_\mathrm{dust}$.
Equations (\ref{eq:dust_temp}) and (\ref{eq:dustsed}) indicate
that we determine the FIR luminosity so that
it is consistent with the observed 880 $\micron$ flux
as explained at the end of this paragraph
(see also Fig.\ 4 in H11).
In H11, we gave $R_\mathrm{dust}$
as a free parameter, but in this paper, we determine
the typical radius for the distribution of the dust
associated with the central star-forming region,
to which the SMA data is sensitive. We provide
$R_\mathrm{dust}$ by the intensity-weighted
radius as
\begin{eqnarray}
R_\mathrm{dust}^2=\int\!\!\!\int
|\bmath{r}-\bar{\bmath{r}}|^2I(x,\, y)\,
\mathrm{d}x\,\mathrm{d}y\left/
\int\!\!\!\int I(x,\, y)\,\mathrm{d}x\,\mathrm{d}y\right.
,
\end{eqnarray}
where $\bmath{r}=(x,\, y)$ is the projected position
in the image and
$\bar{\bmath{r}}$ is the intensity-weighted centre of
the image.
Although our 880 $\micron$ images are contaminated with
free--free emission, \citet{beck01} show similar
spatial distributions between mid-infrared dust emission
and radio free--free emission in He 2-10, implying that
it is reasonable to assume that both dust and free--free
have similar spatial extent at 880 $\micron$ at least
for this galaxy. For NGC 5253, \citet{turner04} show
that most of the free--free flux comes from the central
1.2 arcsec region, which implies that dust emission is more
extended than free--free emission. In such a case,
$R_\mathrm{dust}$ underestimates the actual extension
of dust. The value of
$R_\mathrm{dust}$ for each galaxy is listed in
Table~\ref{tab:quantities}. After all, with
$R_\mathrm{dust}$ (given in Table \ref{tab:quantities}),
$M_*$ (derived from the free--free luminosity
through equation \ref{eq:ff})
and $\mathcal{L}_\mathrm{OB}$
(estimated by equation \ref{eq:LOB}), the unknown
parameters are $M_\mathrm{dust}$ and
$T_\mathrm{dust}$ (note that $\tau_\mathrm{dust}$
in equation \ref{eq:dust_temp} is given by
equation \ref{eq:taudust}). These two unknowns are
obtained by solving
equations (\ref{eq:dust_temp}) and (\ref{eq:dustsed}).

\begin{table*}
\centering
\begin{minipage}{150mm}
\caption{Derived or estimated quantities.}
\label{tab:quantities}
\begin{tabular}{lccccccccc}
\hline
Object & \multicolumn{2}{c}{$R_\mathrm{dust}$} &
$f_\mathrm{dust}\,^\mathrm{a}$ &
$f_\nu (15~\mathrm{GHz})$ & $M_*$ & $M_\mathrm{dust}$ &
$T_\mathrm{dust}$ & $\tau_\mathrm{dust}$ & $L_\mathrm{FIR}$\\
 & (arcsec) & (pc) & (mJy) & (mJy)& ($10^6~\mathrm{M}_{\sun}$) &
($10^4~\mathrm{M}_{\sun}$) & (K) & & ($10^9~\mathrm{L}_{\sun}$)\\
\hline
He 2-10  & 3.31 & 168 & 9--16 & 21.1--17.5 & 11--9.4 &
4.7--9.2 & 48--46 & 0.69--1.4 & 3.2--4.0\\
NGC 5253 & 3.61 & 64.8 & 12--26 & $65\pm 6\,^\mathrm{b}$ & $4.3\pm 0.4$ &
0.65--1.5 & 57--53 & 0.64--1.5 & 1.2--1.9\\
II Zw 40 & 3.28 & 167 & $<5.6$ & 14--18.5 & $7.5\pm 0.8$ &
$<3.7$ & 45$^\mathrm{c}$ & $<0.55$ & $<1.8$ \\
\hline
\end{tabular}

\medskip

$^\mathrm{a}$ The dust flux estimated from the total SMA
880 $\micron$ flux minus the free--free contribution
(Table \ref{tab:data}).\\
$^\mathrm{b}$ Converted from the 3.1 mm flux ($54\pm 5$ mJy) in
\citet{meier02} by assuming frequency dependence $\nu^{-0.1}$.\\
$^\mathrm{c}$ The value is for the upper limit of $L_\mathrm{FIR}$.
\end{minipage}
\end{table*}

\subsection{Derived quantities}\label{subsec:quantities}

The quantities derived by the models above
($M_*$, $M_\mathrm{dust}$, and $T_\mathrm{dust}$)
are listed in Table \ref{tab:quantities}. The stellar
mass formed by the current starburst ranges from
a few $\times 10^6$ to $\sim 10^7$ M$_{\sun}$, and
the dust mass associated with the star-forming regions is
$\sim 10^4$--$10^5$ M$_{\sun}$. The solar-metallicity
sample, He 2-10, has the largest dust content: if the
dust mass is normalized to the stellar mass, its
dust-to-stellar mass ratio is 4--10 $\times 10^{-3}$
in comparison with the values 2--3 $\times 10^{-3}$
for NGC 5253 and $<5\times 10^{-3}$ for II Zw 40.
This indicates that He 2-10 is the most dust-enriched
system probably
because of the highest metallicity. As argued in H11
(see also Section \ref{subsec:dust_prop}),
the dust is preexisting or grown in dense molecular
clouds. Since dust growth is efficient in high-metallicity
environments, the highest dust content in the He 2-10 centre
among the three sample BCDs can be interpreted to be
the consequence of the most
efficient dust growth. Dust temperatures ($\sim 50$--60~K)
higher than those in the Milky Way ($\sim 15$--20~K;
\citealt{draine84}) are obtained for all the
sample, supporting that intense star formation is occurring
in compact regions. {Direct constraints on
$T_\mathrm{dust}$ from FIR data are crucial in future
high-resolution FIR data around the spectral peak,
since $L_\mathrm{FIR}$ is
the most sensitive to $T_\mathrm{dust}$ than to the
other parameters (see equation \ref{eq:dust_temp}).
However, because $M_\mathrm{dust}$ works as an
adjusting factor under a given 880 $\micron$ flux
through equation (\ref{eq:dustsed}), the uncertainty in
$L_\mathrm{FIR}$ is not so large as expected from
the uncertainty in $T_\mathrm{dust}$.
}

\subsection{Radio--FIR relation}\label{subsec:radio_FIR}

In Section \ref{sec:obs}, we have shown that a large
fraction of the SMA 880~$\micron$ flux in the sample BCDs
is contributed from free--free emission,
while the submm emission is usually dominated
by dust on global scales of galaxies \citep{galliano05}.
{The large contribution from free--free emission
to the submm emission may cause a significant impact
on the radio--FIR relation in such a way that
the radio luminosity is relatively enhanced (H11).}
Now we examine this
issue by plotting the radio--FIR relation for the
central regions in our sample BCDs.

\begin{figure}
\includegraphics[width=0.45\textwidth]{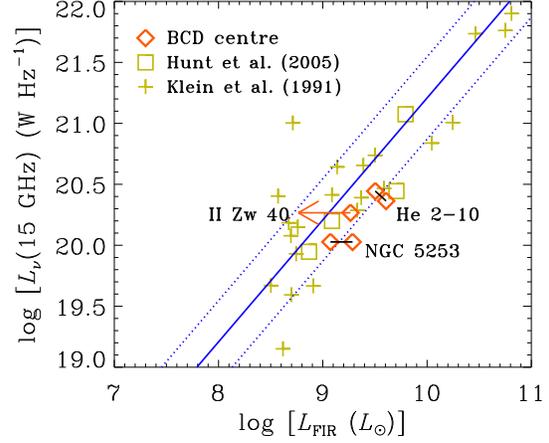}
\caption{Radio--FIR relation for the central star-forming
regions (the diamonds connected by
the solid lines for He 2-10 and NGC 5253 and the
diamond with an arrow for II Zw 40), in terms of the
global relations for BCDs (squares and crosses).
The observational data for the global emission from
BCDs are taken from \citet{hunt05} (squares) and
\citet{klein91} (crosses).
We also show the solid line with $q_{15}=2.80$
and the dotted lines with $q_{15}=2.80\pm 0.34$
(i.e.\ $\pm 1~\sigma$).
\label{fig:radio_fir}}
\end{figure}

In Fig.\ \ref{fig:radio_fir}, we show the relation
between the monochromatic luminosity at 15 GHz
[$L_\nu (15~\mathrm{GHz})$] and the FIR luminosity
$L_\mathrm{FIR}$ for the central star-forming regions
in the sample BCDs. $L_\mathrm{FIR}$ has already been
obtained in Section \ref{subsec:quantities} (see also
Table \ref{tab:quantities}),
while $L_\nu (15~\mathrm{GHz})$ is obtained by using
the flux at 15 GHz in Table \ref{tab:quantities} multiplied by
$4\pi D^2$ (see also
Sections \ref{subsec:he2-10}--\ref{subsec:iizw40}).
The upper and lower bounds of $L_\mathrm{FIR}$
correspond to the lower and upper bounds of
$f_\mathrm{dust}$ (the dust flux at 880 $\micron$)
given in Table \ref{tab:quantities}.
{(Recall that $f_\mathrm{dust}$ is the
total SMA 880 $\micron$ flux minus the free--free
contribution. If the free--free contribution were
not subtracted, we would overestimate $L_\mathrm{FIR}$
by a factor of 2 or more.)}
For He 2-10, the upper and lower limits of
$L_\nu (\mathrm{15~GHz})$ are used,
for NGC~5253, we use the measured values, and
for II Zw 40, we plot the lower value for
$L_\nu (\mathrm{15~GHz})$ to put an upper limit for
$L_\mathrm{FIR}$ (Fig.\ \ref{fig:radio_fir}).

For comparison, we plot the observational data
of global emission from BCDs (i.e.\ the total
luminosity from the entire system) in
Fig.\ \ref{fig:radio_fir}. The sample is
taken from \citet*{hunt05} for the FIR and 15~GHz
global luminosities (squares) and
\citet*{klein91} for the FIR and 10.7 GHz global
luminosities (crosses). The 10.7 GHz luminosity is
converted to the 15 GHz luminosity by assuming
the frequency dependence to be
$\propto\nu^{\langle\alpha\rangle}$, where the spectral
index $\langle a\rangle$ is given in \citet{klein91}
as a result of fitting to the data at some available
radio frequencies (even if we assume
$\langle\alpha\rangle =-0.1$ for all the sample,
the change of
15 GHz luminosity is too slight to affect our
results). As an observational estimate of
the FIR luminosity, we adopt an empirically
derived formula by \citet{nagata02}, who estimate
the total dust
luminosity at $\lambda\geq 40~\micron$
by using the \textit{IRAS} 60 and 100 $\micron$
fluxes.\footnote{The contribution at 8--40 $\micron$
is neglected since the contribution from these
wavelengths to the total FIR is small
(at most 30 per cent).
{The contribution at
$\lambda >120~\micron$ is included, while it is not
included in the FIR luminosity defined by \citet{helou85}.}}
The data are summarized in
Appendix \ref{app:radio_FIR}.

We adopt the radio-to-FIR ratio
as usually used \citep[e.g.][]{condon92}. Here
we define $q_{15}$ as
\begin{eqnarray}
q_{15} & \equiv & \log\left(
\frac{L_\mathrm{FIR}}{3.75\times 10^{12}~\mathrm{W}}
\right)
-\log\left[
\frac{L_\nu (15~\mathrm{GHz})}{\mathrm{W~Hz}^{-1}}
\right] .
\end{eqnarray}
The average of $q_{15}$ for the global luminosities
is $q_{15}=2.80$ with a standard deviation of 0.34.
We show the lines with constant $q_{15}$ in
Fig.\ \ref{fig:radio_fir}. We observe that
the data points for the central star-forming regions
in our sample are consistent with the range of $q_{15}$
(i.e.\ $2.80\pm 0.34$) that explains
the global radio--FIR relation of BCDs.

Naively, it would be expected that low-metallicity
galaxies have relatively small amount of dust,
so that the stellar emission may not be efficiently
reprocessed into FIR \citep{hirashita_hunt08}.
On the other hand,
free--free emission does not have such a
dependence on metallicity (or dust abundance).
Thus, we would expect systematically smaller $q_{15}$
for low-metallicity galaxies. However, NGC~5253
[$12+\log (\mathrm{O/H})=8.14$],
compared with He 2-10 [$12+\log (\mathrm{O/H})=8.93$],
does not follow this expectation, which means that
even a low metallicity environment can reprocess
the stellar light into FIR with a similar
efficiency to a solar metallicity environment.
As we can see in Table \ref{tab:quantities}, the
dust optical depth is
comparable between He 2-10 and NGC~5253, although the
dust mass is smaller in NGC 5253 than in He 2-10.
This is because the distribution of dust is
more compact in NGC 5253 than in He 2-10. A compact
geometry of dust distribution tends to predict a
high dust temperature \citep[e.g.][]{takeuchi05}, explaining
the observed higher dust temperature in NGC 5253 than in
He 2-10.
{The radio--FIR relation of II Zw 40 is still
uncertain because of the uncertainty in $L_\mathrm{FIR}$.
However, it is still possible that this galaxy has a similar
$q_{15}$ to the other two galaxies. If its
$q_{15}$ is significantly smaller than the other
two galaxies, we need to consider the reason other than
the metallicity, since NGC 5253, which has a similar
metallicity to II Zw 40, has a $q_{15}$ value as large
as that of a solar metallicity object, He 2-10.}

The radio--FIR relation of star-forming
dwarf galaxies is similar to that of normal
galaxies in spite of the difference
in metallicity (\citealt{klein91};
\citealt*{hopkins02}; \citealt{wu08}), although
some specific galaxies show deviations
\citep{cannon06}. Thus, metallicity
(or dust content) cannot be the dominant factor that
governs the radio--FIR relation.
\citet{cannon05} find a spatial variation
of FIR-to-radio ratio by an order of magnitude
in a metal poor dwarf galaxy, IC 2574.
\citet{dumas11} also
show different radio--FIR relations between
spiral arms and interarm regions in M51.
These observations also support the above
statement that
metallicity is not the dominant factor that
varies the FIR-to-radio ratio.
{
The robustness of the FIR-to-radio ratio among our
BCD sample is further discussed and interpreted
in Section \ref{subsec:ff_contribution}.
}


\subsection{Molecular gas}\label{subsec:co}

Gas mass provides the normalization in estimating
star formation efficiency and dust-to-gas ratio.
The observed wavelength range also covers
the CO(3--2) rotational
transition line at 345.796 GHz (at rest).
Although quite a lot of detections are reported for
CO in He 2-10
\citep[e.g.][]{baas94,kobulnicky95,vanzi09}, we use our
data which have the advantage of covering the
the same $(u,\, v)$ range. In Table \ref{tab:co},
we summarize the
quantities derived by our CO(3--2) data.

\begin{figure}
\includegraphics[width=0.45\textwidth]{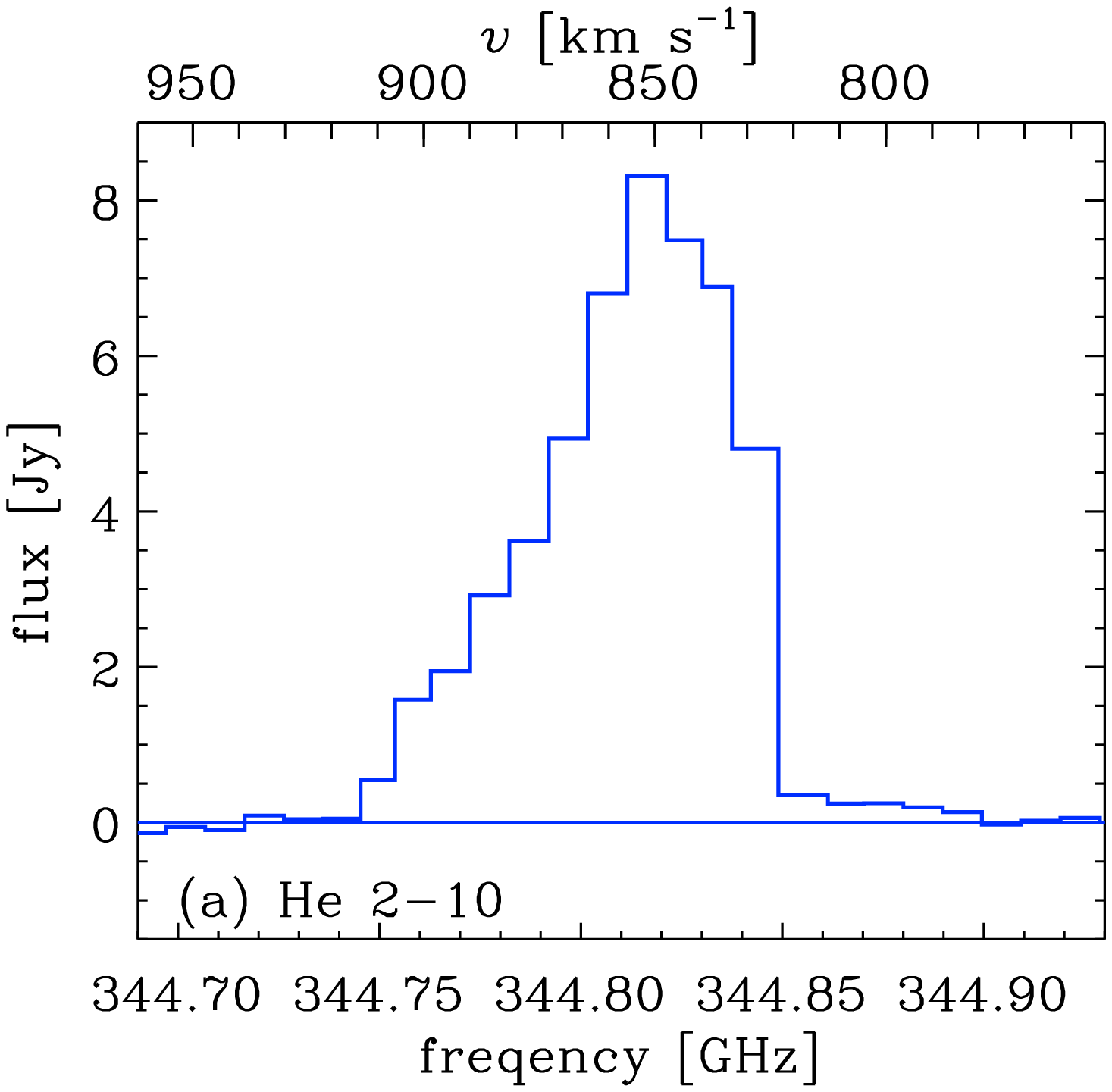}
\includegraphics[width=0.45\textwidth]{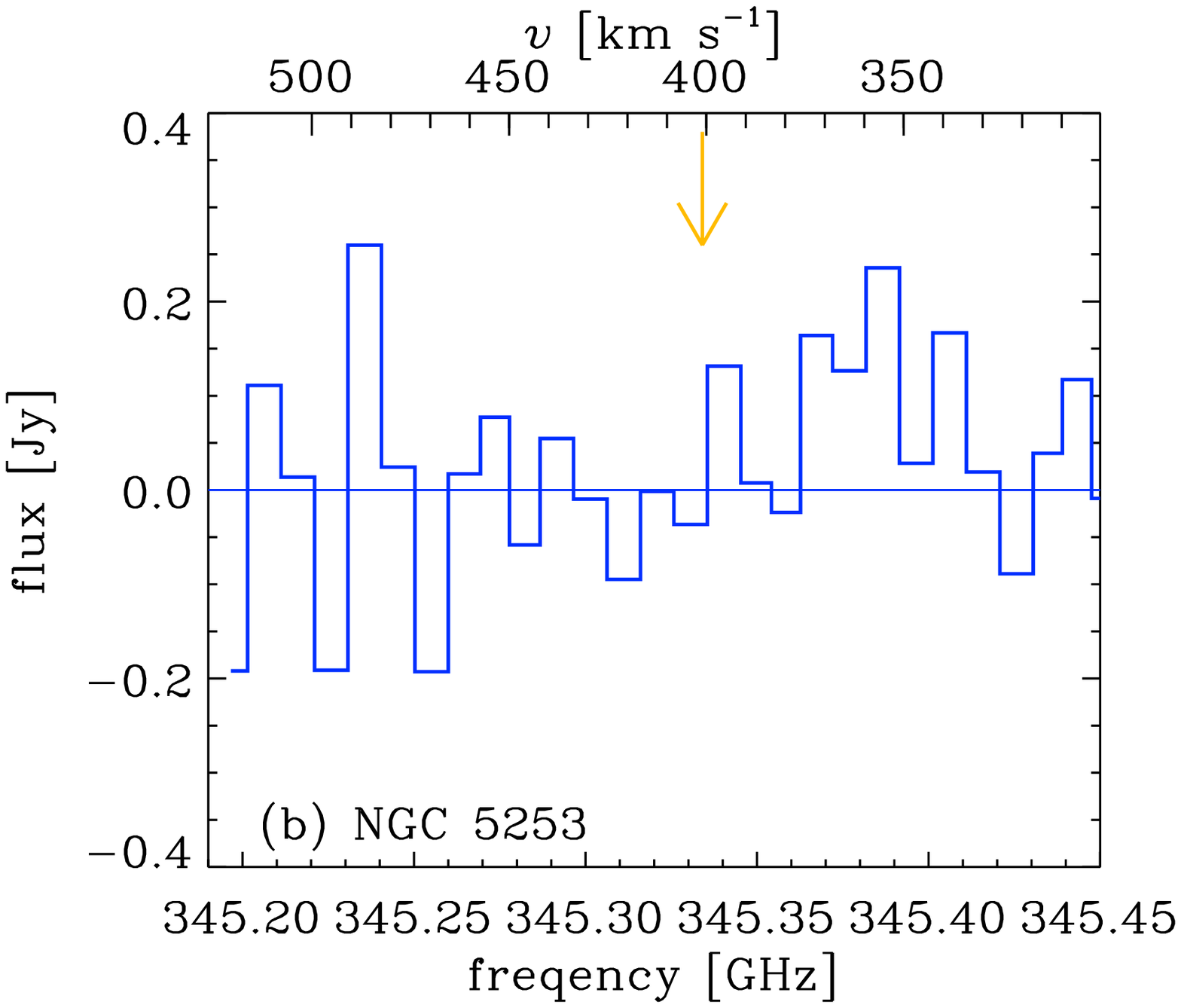}
\includegraphics[width=0.45\textwidth]{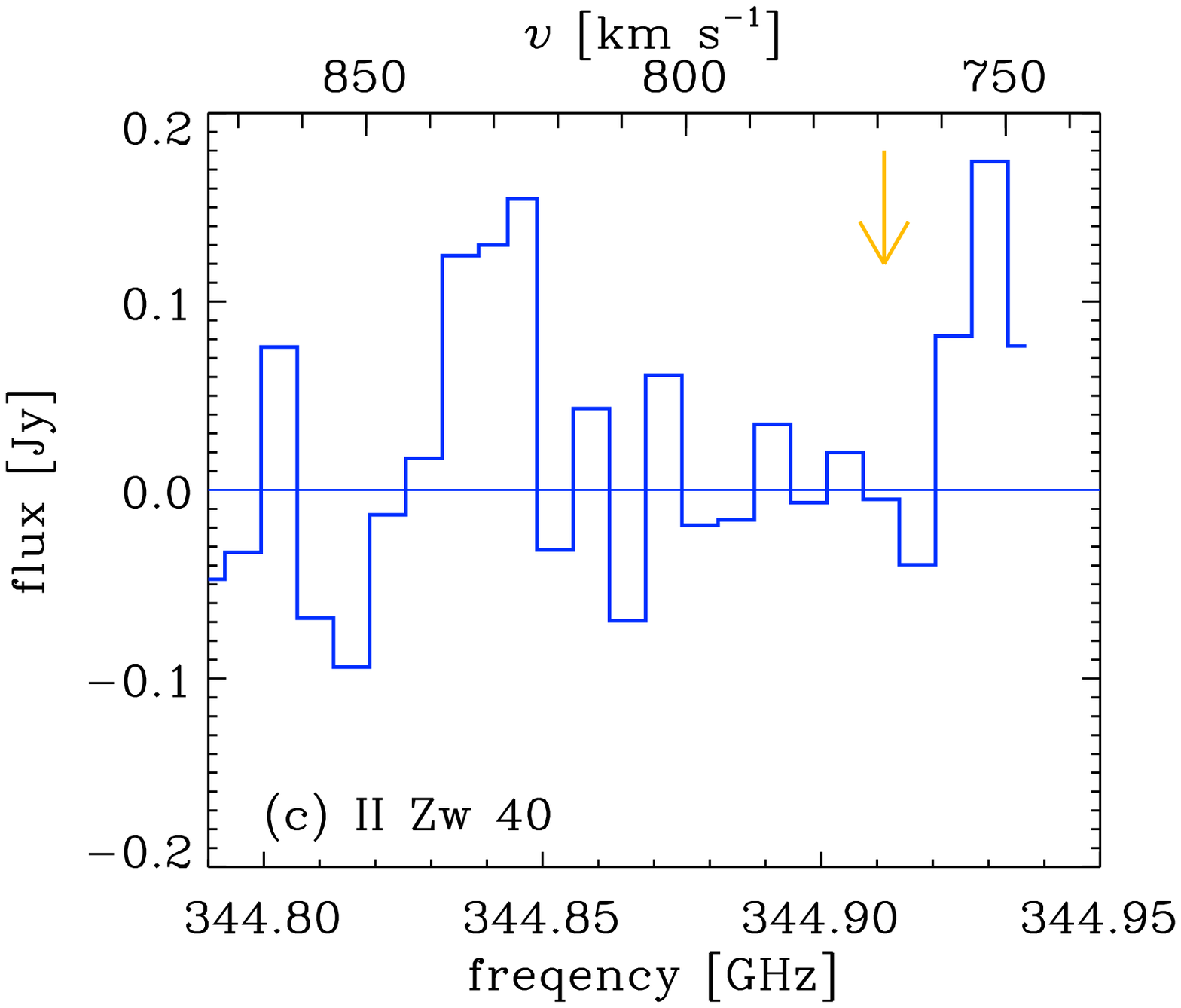}
\caption{
Spectra around the frequency where the
CO(3--2) line is expected. Panels (a), (b), and (c)
show the spectrum in the central part of
He 2-10, NGC~5253, and II Zw 40, respectively.
The LSR velocity is also shown on the upper axis.
For the latter two galaxies, we did not detect
significant CO(3--2) emission, whose expected
frequency is marked by the arrow.
\label{fig:co}}
\end{figure}

The spectra around the expected wavelength of CO(3--2)
are shown in Fig.\ \ref{fig:co}. The emission is only
detected in He 2-10. For II Zw 40, we failed to obtain
the data at $>344.94$ GHz. Yet, we can conclude that
CO(3--2) is not detected in this galaxy at the
expected frequency.

For He 2-10, we derive the CO(3--2) line flux
($S_\mathrm{CO(3-2)}$), the FWHM of the line
in units of velocity ($\Delta V$), and the radius
of the CO emitting region [$R=0.7(ab)^{1/2}$,
where $a$ and $b$ are respectively the FWHM major
and minor axes; \citet{meier02}].
For NGC 5253 and II Zw 40, we derive upper limits
for the CO(3--2) line flux. The total mass
(dynamical mass) in the central region traced by SMA
can be estimated
by \citep{maclaren88,meier02}
\begin{eqnarray}
M_\mathrm{tot}=189\left(
\frac{\Delta V}{\mathrm{km~s}^{-1}}\right)^2\left(
\frac{R}{\mathrm{pc}}\right)~\mathrm{M}_{\sun}.
\end{eqnarray}
The molecular mass denoted as $M_\mathrm{mol}$ is
estimated as \citep{meier01}
\begin{eqnarray}
M_\mathrm{mol} & = & 1.23\times 10^4\left(
\frac{X_\mathrm{CO}}{X_\mathrm{COgal}}\right)
\left(\frac{115~\mathrm{GHz}}{\nu}\right)^2
\left(\frac{D}{1~\mathrm{Mpc}}\right)^2\nonumber\\
& & \times \left(
\frac{S_\mathrm{CO(3-2)}}{\mathrm{Jy~km~s}^{-1}}
\right)R_{32/10}^{-1},
\end{eqnarray}
where $X_\mathrm{COgal}=2.3\times 10^{20}~
\mathrm{cm^{-2}~(K~km~s^{-1})^{-1}}$ is the
Galactic conversion factor \citep{strong88},
$X_\mathrm{CO}$ is the metallicity-dependent
conversion factor \citep{arimoto96}, and $R_{32/10}$
is the CO(3--2)/CO(1--0) line ratio. We adopt
$R_{32/10}=0.6$ for all the sample
as a representative value for dwarf starbursts
\citep{meier01}.
The molecular gas mass is compared to
the dust mass and the stellar mass to obtain,
respectively,
the dust-to-gas mass ratio
($\mathcal{D}\equiv M_\mathrm{dust}/M_\mathrm{mol}$),
and $\epsilon_*$
[$\epsilon_*\equiv M_*/(M_\mathrm{mol}+M_*)$],
which under normal conditions (in particular
no molecule destruction), provides the star-formation
efficiency.

\begin{table*}
\centering
\begin{minipage}{170mm}
\caption{Molecular gas properties in the center of
the sample BCDs.}
\label{tab:co}
\begin{tabular}{lccccccccc}
\hline
Object & $S_\mathrm{CO(3-2)}$ & $\Delta V$ &
\multicolumn{2}{c}{$R$} & $M_\mathrm{tot}$ &
$M_\mathrm{mol}$ & $X_\mathrm{CO}/X_\mathrm{COgal}$ &
$\epsilon_*$ & $\mathcal{D}$ \\
 & (Jy km s$^{-1}$) & (km s$^{-1}$) & (arcsec) & (pc)&
($10^6~\mathrm{M}_{\sun}$) & ($10^6~\mathrm{M}_{\sun}$) & &
& ($10^{-3}$)\\
\hline
He 2-10  & $387\pm 8$ & $40.8\pm 1.4$ & 6.48 & 330 &
$100\pm 8$ & $94\pm 9$ & 0.97 & 0.08--0.11 & 0.50--0.98\\
NGC 5253 & $<20^\mathrm{a}$ & --- & --- & ---
& --- & $<3.8$ & 6.0 & $>0.53$ & $>1.7$ \\
II Zw 40 & $<13^\mathrm{a}$ & --- & --- & --- & --- & $<20$
& 6.2 & $>0.27$ & $>0.27$\\
\hline
\end{tabular}

\medskip

$^\mathrm{a}$ 3 $\sigma$ upper limits with
$\Delta V=40$ km s$^{-1}$.
\end{minipage}
\end{table*}

We detected CO(3--2) only in He 2-10, while we obtained
only upper limits for NGC 5253 and II Zw 40.
Our observations are sensitive only to the centrally
concentrated component. Diffuse components, if any, may
be resolved out.
{Table \ref{tab:co} indicates that
$(M_\mathrm{mol},\, M_\mathrm{tot})=
(9.4\times 10^7~\mathrm{M}_{\sun},\, 1.0\times 10^8~
\mathrm{M}_{\sun})$ for the central star-forming
region of He 2-10 with $\Delta V=40.8$ km s$^{-1}$.}
\citet{vanzi09} reported
$\Delta V=53$ km s$^{-1}$ for the central 20 arcsec
in He 2-10.
\citet{meier01} estimate
$(M_\mathrm{mol},\, M_\mathrm{tot})=
(1.4\times 10^8~\mathrm{M}_{\sun},\, 3.3\times 10^8~
\mathrm{M}_{\sun})$ and
$(3.7\times 10^7~\mathrm{M}_{\sun},\,
9.4\times 10^7~\mathrm{M}_{\sun})$ for He 2-10 and NGC 5253,
respectively. Their large velocities and masses are due to
different spatial scales traced.
No CO(3--2) detection has been reported
for II Zw 40, but the molecular mass derived from CO(2--1)
and CO(1--0) ($\ga 5\times 10^6$ M$_{\sun}$) is
not contradictory with our upper limit.

In the He 2-10 centre, $M_\mathrm{tot}\simeq M_\mathrm{mol}$,
and the stellar mass only occupies a small fraction
($\sim 10$ per cent) of the total mass.
A star formation efficiency of $\sim 10$ per cent is
near the values derived for the Galactic giant molecular
clouds \citep*{lada10}. The dust-to-gas ratio is significantly
smaller than the Galactic value
($\sim 6\times 10^{-3}$; \citealt{spitzer78}) although the
metallicity is similar. This is probably because we only
trace the high-temperature dust component that is
directly heated by the
current starburst without being shielded by other
dust components. A similar underestimate of dust
mass is also reported by \citet{sun11}.
Thus, we may miss a large amount of
dust if a small amount of dust efficiently shields the
stellar light.

For the low-metallicity objects, NGC 5253 and
II Zw 40, the upper limits of molecular gas mass
can be used to constrain the lower limits for
$\epsilon_*$ and $\mathcal{D}$. Both objects have
significantly larger $\epsilon_*$ than He 2-10. This
is interpreted in two ways: (i) the star formation
efficiency is actually high, or (ii) molecular
gas is dissociated quickly ($\la 3$ Myr), causing
an underestimate of the total gas mass.
These two possibilities are further discussed in
Section \ref{subsec:gas}.

\section{Discussion}\label{sec:discussion}

\subsection{Free--free contribution at 880 $\micron$}
\label{subsec:ff_contribution}

In H11, the 880 $\micron$ luminosity in the II Zw 40
centre is shown to be dominated by free--free
emission. This is why H11 suggests that free--free
dominated emission at $\sim 880~\micron$
can be used to select extremely young starbursts
($\la 3$ Myr). In this paper, this is confirmed
in the sense that the
central star-forming regions in He 2-10 and NGC 5253
also have significant contributions of free--free
emission at 880~$\micron$
(Section \ref{sec:obs}). {There is a hint} in
Table \ref{tab:data} that as metallicity becomes higher
the free--free fraction becomes lower (i.e.\
the fraction of dust emission becomes higher).
{This trend should be checked with a larger
sample in the future.}

Although free--free emission has a significant
contribution even at 880~$\micron$ [while global
880~$\micron$  emission {(`global' refers to
the integrated emission of the entire galaxy)} is generally
dominated by dust \citep[e.g.][]{hunt05}],
the radio--FIR relation of the central {regions}
in our sample BCDs is surprisingly
consistent with that defined by using global
luminosities as shown in Fig.\ \ref{fig:radio_fir}.
This is probably due to the high dust temperatures,
which enhance the FIR luminosity and compensates the
small contribution from dust emission at 880~$\micron$.
Therefore, it is possible that this
compensating effect
is a key to understand the robustness of the
radio--FIR relation.

The robustness of the radio--FIR relation may
simply reflect the fact that both radio and
FIR luminosities are good
indicators of star formation activities \citep{condon92}.
If we \textit{assume} the robustness of the
radio--FIR relation, we can explain the small
contribution of dust emission at 880 $\micron$ in
comparison with free--free emission as follows.
By assumption, the free--free flux at 880 $\micron$
[$f_\mathrm{ff}(\nu )$] is just
proportional to the total FIR flux, which is
proportional to $T_\mathrm{dust}^6$
(for dust mass absorption coefficient $\propto \nu^2$;
\citealt{draine84}).
On the other hand, if the Rayleigh--Jeans
approximation is valid, the 880 $\micron$ dust flux,
$f_\mathrm{dust}(\nu )$, is proportional to
$T_\mathrm{dust}$. Therefore,
$f_\mathrm{ff}(\nu )/f_\mathrm{dust}(\nu )\propto
T_\mathrm{dust}^5$, which indicates that the
fraction of free--free emission at 880 $\micron$
is larger for a higher dust temperature.

\begin{figure}
\includegraphics[width=0.45\textwidth]{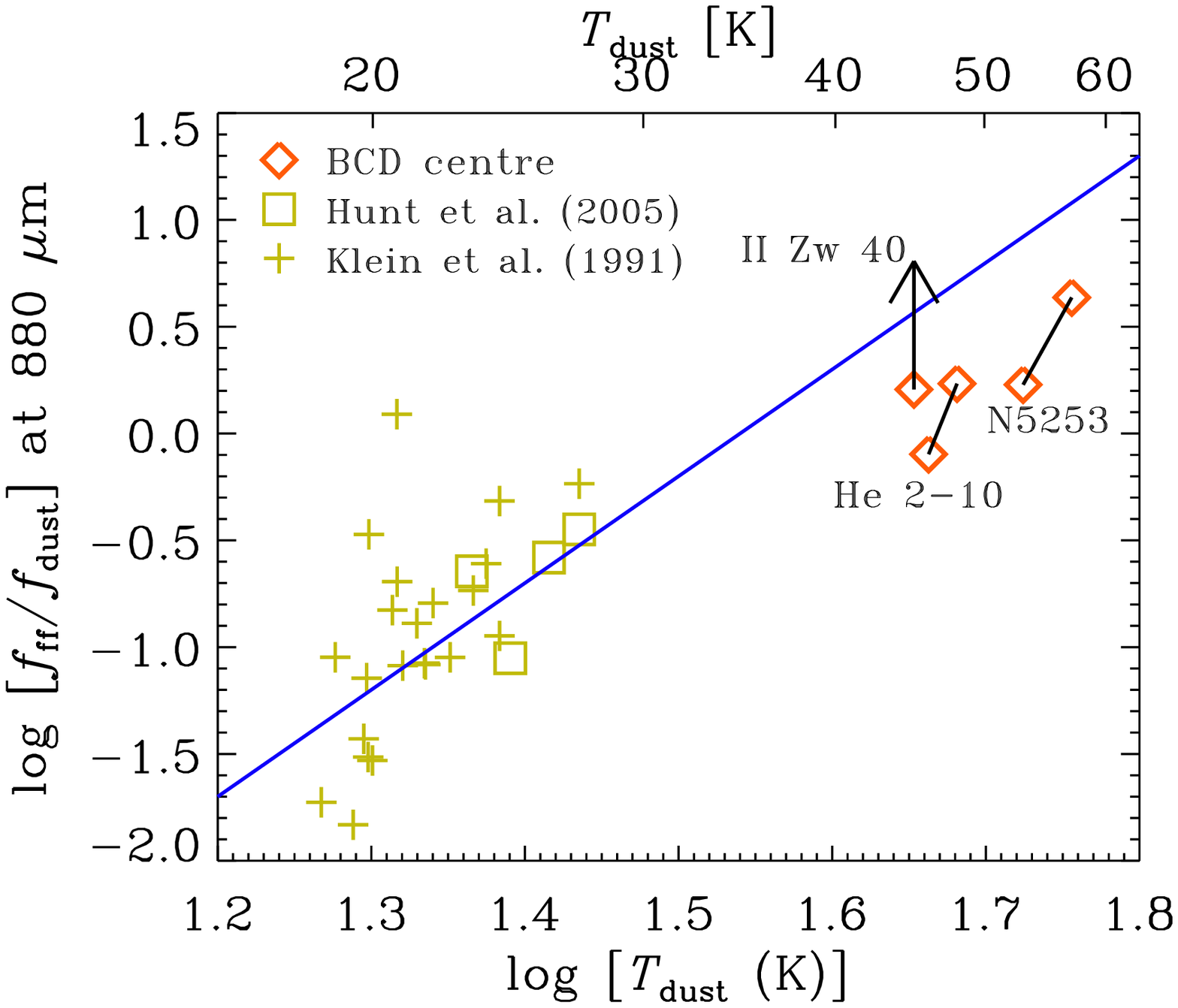}
\caption{Relation between the free--free-to-dust
flux ratio estimated at 880 $\mu$m and the dust
temperature.
He 2-10, II Zw 40, and NGC 5253, in terms of the
global relations for BCDs.
The observational data for the global emission of
BCDs are taken from \citet{hunt05} and \citet{klein91}
for squares and crosses, respectively.
We also show the relation with
$f_\mathrm{ff}/f_\mathrm{dust}\propto CT_\mathrm{dust}^5$,
where a constant $C$ is chosen arbitrarily.
\label{fig:ff_fraction}}
\end{figure}

To examine this temperature dependence of the
free--free fraction, we plot
$f_\mathrm{ff}(\nu )/f_\mathrm{dust}(\nu )$ at
880 $\micron$ in terms of $T_\mathrm{dust}$
for the global emission in the BCDs which were used
to plot the radio--FIR relation
in Fig.\ \ref{fig:radio_fir}
\citep[i.e.\ the samples taken from][]{hunt05,klein91}.
For these BCDs, we estimate $f_\mathrm{dust}$ at
880 $\micron$
as described in Appendix~\ref{app:radio_FIR}.
The free--free flux at 880 $\micron$ ($f_\mathrm{ff}$)
is estimated by converting the free--free
flux at 15 GHz used in Fig.\ \ref{fig:radio_fir}
to that at 880 $\micron$ by assuming a
frequency dependence of $\nu^{-0.1}$.
We also plot our SMA sample for the central parts in
BCDs. We find that there is a positive trend of
$f_\mathrm{ff}/f_\mathrm{dust}$ with
$T_\mathrm{dust}$ as expected above. We also
draw a line with
$f_\mathrm{ff}/f_\mathrm{dust}\propto T_\mathrm{dust}^5$
with an arbitrary normalization in
Fig.\ \ref{fig:ff_fraction}.
This line roughly bridges our sample and the
above samples, although the large scatter of the
data implies that our discussion may be too simplified.
Supported by the correlation between
$f_\mathrm{ff}/f_\mathrm{dust}$ and $T_\mathrm{dust}$,
we conclude that the
large contribution of free--free emission in
our observation is a natural consequence of
picking up active star-forming regions which
have high dust temperatures.

{The large free--free fraction in the central
star-forming regions of our sample indicates that
the subtraction of free--free
emission at submillimetre wavelengths is a critical
step in studying the dust emission.
\citet{galliano05} find that the submm excess
in some BCDs
needs to be fitted with `very cold grains',
whose derived abundance may be sensitive to the
subtraction of
free--free emission. Submm excess is preferentially
seen in low-metallicity dwarf galaxies \citep{galametz11}.
Since we have assumed that the SMA 880 $\micron$ flux
is dominated by the
warm dust component heated by young stars,
we cannot test the existence of
this very cold component in the central star-forming
regions. We need to wait for multi-wavelength observations
at FIR--submm wavelengths with spatial
resolutions of a few arcsec to tackle this issue.
}

\subsection{Dust abundance}\label{subsec:dust_prop}

As mentioned in Section \ref{subsec:quantities},
the dust-to-stellar mass ratio is larger in He 2-10
than in the other two galaxies, while as
shown in Section~\ref{subsec:co},
the dust-to-molecular gas mass ratio does not
necessarily trace the trend of metallicity.
The dust-to-gas ratio derived for He 2-10 is
much lower than the Galactic dust-to-gas ratio
($6\times 10^{-3}$; \citealt{spitzer78}),
although He 2-10 is a solar-metallicity object.
As interpreted in Section \ref{subsec:co},
the dust mass in He 2-10 may be underestimated
because we only trace the dust directly heated
by the central star clusters; that is, the dust mass
estimate from dust emission is not
sensitive to the dust in regions where the UV radiation
from the central stars is shielded.

As argued in H11, the observed dust should
either be preexisting or have grown by accretion
in the dense star-forming regions. According to
\citet{hirashita_kuo}
\citep[see also][]{inoue11,asano12}, the dust-growth
time-scale for silicate (a similar time-scale is
obtained for carbonaceous dust) is estimated as
\begin{eqnarray}
\tau_\mathrm{grow} & \simeq & 2.1\times 10^5~
\mathrm{yr}\left(
\frac{\langle a^3\rangle /\langle a^2\rangle}{0.1~\micron}
\right)
\left(\frac{Z}{1~\mathrm{Z}_{\sun}}\right)^{-1}\nonumber\\
& &
\times\left(\frac{n_\mathrm{H}}{10^5~\mathrm{cm}^{-3}
}\right)^{-1}
\left(\frac{T_\mathrm{gas}}{50~\mathrm{K}}\right)^{-1/2}
\left(\frac{S}{0.3}\right)^{-1}\, ,
\end{eqnarray}
where $\langle a^3\rangle$ and $\langle a^2\rangle$
are the averages of $a^3$ and $a^2$ ($a$ is the
grain radius) for the grain size distribution
(we adopt
$\langle a^3\rangle /\langle a^2\rangle =0.1~\micron$),
$Z$ is the metallicity
(we adopt $Z=1.73$, 0.28, and 0.28~Z$_{\sun}$ for
He 2-10, NGC 5253, and II Zw 40, respectively, by assuming
that the solar oxygen abundance is
$12+\log (\mathrm{O/H})=8.69$; \citealt{lodders03}),
$n_\mathrm{H}$ is the
hydrogen number density (we adopt
$n_\mathrm{H}=10^5$ cm$^{-3}$ for the dense star-forming
regions in BCDs; H11), $T_\mathrm{gas}$ is the gas
temperature (we adopt $T_\mathrm{gas}=50$ K;
\citealt{wilson97}) and $S$ is the sticking efficiency of
the relevant metal species onto the dust surface
(we adopt $S=0.3$; \citealt{leitch85,grassi11}).
Then, we obtain
$\tau_\mathrm{grow}\sim 0.12$, 0.75 and 0.75 Myr
for He 2-10, NGC 5253, and II Zw 40, respectively.
These time-scales are shorter than or comparable to the
ages of the star-forming regions ($\la$ a few Myr).
Therefore, dust growth by the accretion of gas-phase
metals can be
an effective mechanism of increasing the dust
content in the central star-forming regions of
those BCDs.


\subsection{Molecular gas properties}\label{subsec:gas}

In Section \ref{subsec:co}, we have shown that the
evaluated star formation efficiencies in the central
parts of the low-metallicity BCDs (NGC~5253 and II~Zw~40)
are high. Such high star formation efficiencies
in low-metallicity galaxies are already reported when
CO emission is used for the
molecular gas tracer \citep[e.g.][]{schruba12}. However,
no physical mechanism that makes the star formation
in metal-poor gas efficient is known.
\citet{hirashita02} suggest that star formation is
rather inefficient in low metallicity (i.e.\
low dust-to-gas ratio) environments because of
inefficient H$_2$ formation \citep[see also][]{gnedin11}.
Another interpretation of the high star formation
efficiencies is that CO is more easily dissociated in
low-metallicity environments because of less shielding
of dissociating
photons by dust
{\citep*{wolfire10,krumholz11,shetty11}}.
\cite{leroy11} suggest that CO dissociation effect
becomes dominant below $12+\log\mathrm{(O/H)}=8.2$--8.4,
which is consistent with CO deficiency of NGC 5253 and
II Zw 40. In such a case, CO is not a good tracer
of the total gas mass, and the gas mass estimated from
CO is an underestimate. The high dust-to-gas ratio in
NGC 5253 can be an artifact of an underestimate of the
molecular gas mass.

\begin{figure}
\includegraphics[width=0.45\textwidth]{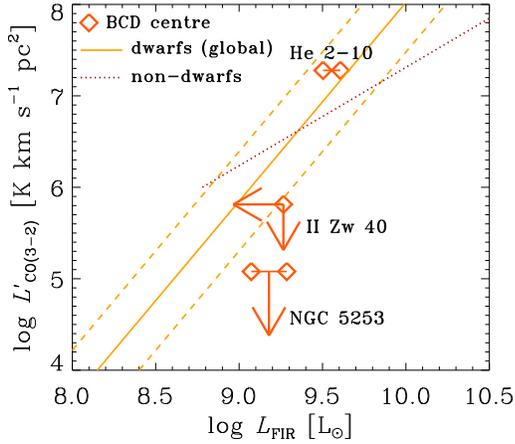}
\caption{Relation between FIR luminosity and
CO(3--2) luminosity for the central part of
He 2-10, NGC~5253, and II Zw 40.
We also show the linear fitting relation
in \citet{meier01} applicable for dwarf galaxies
(solid line) and for non-dwarf galaxies (dotted line;
originally derived from \citealt{mauersberger99}).
The dashed lines show the typical scatter for the
dwarf sample in \citet{meier01}.
\label{fig:fir_co}}
\end{figure}

In order to present the deficiency of CO more clearly,
we show the relation between CO luminosity and
FIR luminosity in Fig.~\ref{fig:fir_co}.
As a reference, we also show the global
(i.e.\ entire galaxy) relation of a dwarf sample
in \citet{meier01}
(originally from \citealt{mauersberger99}).
Compared with this relation (solid line),
our SMA data point for NGC 5253 is deviated downwards.
This indicates that CO is depleted in the centre
while the FIR dust emission still has a significant
intensity there. The existence of the central
dust emission
demonstrates the existence of gas since gas and dust are
generally well mixed and coexisting;
thus, the natural interpretation of the deficient CO
emission is that CO is highly
dissociated in these low-metallicity galaxies.
Another low-metallicity galaxy, II~Zw~40, may also
be deviated downward, but since we only obtained
an upper limit for the FIR luminosity, II Zw 40 may
be consistent with the global relation. The
relatively rich molecular gas content in He 2-10
may be explained by a stronger shielding of dissociating
radiation due to its
rich dust content compared with the other two
BCDs. This trend of CO abundance with metallicity is
consistent with
theoretical predictions by \citet{krumholz11}.

We also show in Fig.\ \ref{fig:fir_co} a line
for non-dwarf galaxies (mostly spiral galaxies).
Such a correlation between CO and FIR luminosities
has been well investigated \citep[e.g.][]{gao04}.
The relation for the dwarf sample (solid line)
is systematically deviated.
The deviation is even larger if we focus on
the central part of NGC 5253. It is probably because
we pick up a region with an intense radiation field
(i.e., strong CO dissociation)
if we pick up the central region.

\subsection{Implication for high-redshift
observations}

High-redshift starbursts are often traced by
observations of dust continuum and/or CO
\citep[e.g.][]{michalowski10}.
BCDs, which host compact star-forming regions
with various evolutionary stages
(i.e.\ various metallicities),
provide good `laboratories' of high-$z$ galaxies
(or galaxy evolution), although our sample size
is still limited. For the He 2-10 centre, the gas
mass derived from the CO observation through
a conventional conversion factor matches the total
dynamical mass for the He 2-10 centre, supporting
that CO is a good tracer of the total gas mass
in this particular BCD. However, as
discussed in Section \ref{subsec:co}, large values of
star formation efficiency
in low-metallicity BCDs, NGC 5253 and II Zw 40,
imply that CO luminosity is not necessarily a
good indicator
of the total gas mass in low-metallicity
starburst environments.
Thus, if a high-$z$ galaxy has a
metallicity significantly lower
than solar, CO luminosity has the risk of
underestimating the total gas mass.

This underestimate of gas mass propagates to
the estimate of dust-to-gas ratio.
Moreover, there is a potential risk of
underestimating dust mass because we may miss
low-temperature dust components, which are less
luminous than high-temperature components directly
heated by the current starbursts
(Section \ref{subsec:dust_prop}).
Thus, as long as we trace the dust content by
emission and the gas content by CO lines,
there is a possibility that
the obtained dust-to-gas ratio is not
reliable.

For the origin of dust, it is suggested that
dust growth is the most efficient mechanism of
dust production in high-redshift starbursts
\citep{mattsson11,valiante11}. As suggested in
our BCD observations (Section \ref{subsec:dust_prop}),
high gas densities in intense starbursts
make the grain growth time-scale short.

\section{Conclusion}\label{sec:conclusion}

In order to reveal the submm--radio radiative
properties of young active starbursts in BCDs,
the central star-forming regions in He 2-10,
NGC 5253, and
II Zw 40 were observed in the 340~GHz
(880~$\micron$) band at $\sim 5''$ resolution with
SMA. The 880 $\micron$ fluxes have been
decomposed into free--free and dust emission
components by using centimetre radio data in
the literature. At 880 $\micron$, free--free
emission has proven to have a contribution comparable
to or larger than dust emission in the
central part of all the three BCDs.
We have also shown that the fraction of
free--free emission at 880~$\micron$ has
a positive correlation with the dust temperature.
In spite of the
dominance of free--free emission at 880 $\micron$,
the radio--FIR
relation of the central parts in the BCDs
is consistent with the relation defined for the
global luminosities (i.e.\ the luminosities in
the entire system), supporting the robustness of
the radio--FIR correlation. Finally, we have
analyzed the CO(3--2) emission line, finding that CO
is a good tracer of the total gas mass in He 2-10,
while CO is deficient in the low-metallicity BCDs
(NGC~5253 and II~Zw~40), probably due to
inefficient shielding of dissociating
photons in dust-poor environments.
We also point out that the dust mass
may be potentially underestimated since emission
is always biased to
the dust directly heated by the stars.
Thus, we should keep in mind these potential
underestimates of gas and dust masses in interpreting
submm observations of not only nearby
galaxies but also high-redshift galaxies.

\section*{Acknowledgments}

We thank K. Sakamoto for his continuous help
for the SMA observation and the data analysis,
and S. Beck and S. Matsushita
for helpful discussions on gas and dust emission
properties of galaxies. We thank the
SMA staff for their efforts in running
and maintaining the array. We are grateful to the
anonymous referee for useful comments that improved
this paper very much.
This research has made use of the
NASA/IPAC Extragalactic Database (NED), which
is operated by the Jet Propulsion Laboratory, California
Institute of Technology, under contract with the National
Aeronautics and Space Administration. This research
is supported through NSC grant 99-2112-M-001-006-MY3.

\appendix

\section{Data for the radio--FIR relation}
\label{app:radio_FIR}

We collected radio fluxes around 2 cm (15 GHz) and FIR
(\textit{IRAS} 60 and 100~$\micron$) fluxes of BCDs
in the literature.
We adopt \citet{hunt05} and
\citet{klein91} (Tables \ref{tab:hunt} and \ref{tab:klein},
respectively). The flux listed in these two tables
is a global flux (i.e.\ flux from the entire
galaxy). For the radio fluxes in \citet{hunt05},
if more than two data are available at the same
wavelength, we adopt the data in the lowest-resolution
mode, which is expected to have the smallest missing
flux. We select BCDs with detections both at
\textit{IRAS} 60 and 100 $\micron$, and at 2 cm (15 GHz).
For \citet{klein91}'s sample, we adopt the
Effelsberg 2.8 cm flux and convert it to the 2 cm
flux by adopting average radio spectral indices
$\langle\alpha\rangle$ given in Table 6 of \citet{klein91}.
We exclude BCDs without detection by \textit{IRAS}.
The \textit{IRAS} fluxes of \citet{klein91}'s samples
are taken from \citet{moshir90}.

For these samples, we need to estimate dust
temperatures and fluxes at 880 $\micron$ in
Fig.\ \ref{fig:ff_fraction}. We adopt the following
estimator, $T_\mathrm{LG2}$, for the dust temperature
\citep{nagata02}:
\begin{eqnarray}
T_\mathrm{LG2}=11.8\,
\frac{f_\nu (100~\micron )}{f_\nu (60~\micron )}+
13.8~\mbox{[K]}.
\end{eqnarray}
We use this dust temperature for $T_\mathrm{dust}$
in the text. By using $T_\mathrm{LG2}$,
the dust flux at 880 $\micron$ is estimated as
\begin{eqnarray}
\left. f_\mathrm{dust}(\nu )\right|_{880~\mu\mathrm{m}}
& = & (100/880)^5\,
\frac{\exp (143.9/T_\mathrm{LG2})-1}{\exp(16.35/T_\mathrm{LG2})-1}
\nonumber\\ & & \times f_\nu (100~\micron ).
\end{eqnarray}

\begin{table*}
\centering
\begin{minipage}{100mm}
\caption{\citet{hunt05}'s sample for the radio--FIR
relation.}
\label{tab:hunt}
\begin{tabular}{lcccc}
\hline
Galaxy & $f_\nu (2~\mathrm{cm})$ & $f_\nu (60~\micron)$ &
$f_\nu (100~\micron)$ & $D$ \\
 & (mJy) & (Jy) & (Jy) & (Mpc) \\
\hline
He 2-10  & $21.1\pm 1.2$ & $24.1\pm 2.4$ & $26.4\pm 2.6$ & 10.5\\
NGC 5253 & $54\pm 5$ & $30.5\pm 1.2$ & $29.4\pm 1.8$ & 3.7\\
II Zw 40 & $12\pm 3$ & $6.61\pm 0.70$ & $5.80\pm 0.90$ & 10.5\\
Mrk 33 & $16.0\pm 0.06$ & $4.77\pm 0.04$ & $5.99\pm 0.13$ & 24.9$^\mathrm{a}$\\
\hline
\end{tabular}

\medskip

$^\mathrm{a}$ \citet{tully88}.
\end{minipage}
\end{table*}

\begin{table*}
\centering
\begin{minipage}{120mm}
\caption{\citet{klein91}'s sample for the radio--FIR
relation.}
\label{tab:klein}
\begin{tabular}{lccccc}
\hline
Galaxy$^\mathrm{a}$ & $f_\nu (2.8~\mathrm{cm})$ &
$\langle\alpha\rangle$ & $f_\nu (60~\micron)$ &
$f_\nu (100~\micron)$ & $D$ \\
 & (mJy) & & (Jy) & (Jy) & (Mpc) \\
\hline
Haro 14  & $2.4\pm 0.8$ & $-0.51\pm 0.18\,^\mathrm{b}$ &
  $0.530\pm 0.069$ & $1.04\pm 0.14$ & 13.9 \\
Haro 15 & $6.6\pm 0.5$ & $-0.83\pm 0.21$ &
  $1.35\pm 0.12$ & $1.97\pm 0.20$ & 95.4\\
Mrk 370 & $2.7\pm 0.7$ & $-0.44\pm 0.11$ &
  $1.21\pm 0.11$ & $3.03\pm 0.24$ & 12.9\\
II Zw 40 & $21\pm 2$ & $-0.20\pm 0.05$ &
  $6.61\pm 0.70$ & $5.80\pm 0.90$ & 10.5\\
Haro 1 & $21\pm 3$ & $-0.48\pm 0.08$ &
  $8.57\pm 0.43$ & $12.9\pm 0.65$ & 52.0\\
Mrk 86 & $7.4\pm 2.0$ & $-0.30\pm 0.45$ &
  $3.24\pm 0.19$ & $6.45\pm 0.39$ & 7.0\\
Mrk 401 & $11\pm 4$ & $-0.51\pm 0.18\,^\mathrm{b}$ &
  $2.57\pm 0.13$ & $4.01\pm 0.24$ & 22.2\\
Haro 23 & $5.5\pm 0.8$ & $+0.71\pm 0.30$ &
  $0.401\pm 0.044$ & $0.778\pm 0.132$ & 17.4\\
Mrk 140 & $18\pm 6$ & $-0.29\pm 0.04$ &
  $0.370\pm 0.041$ & $0.630\pm 0.126$ & 22.7\\
Haro 2 & $7.1\pm 0.9$ & $-0.59\pm 0.10$ &
  $4.68\pm 0.28$ & $5.32\pm 0.32$ & 20.5\\
Haro 3 & $9.1\pm 2.0$ & $-0.25\pm 0.02$ &
  $4.95\pm 0.40$ & $6.75\pm 0.41$ & 13.9\\
Mrk 186 & $7\pm 3$ & $+0.44\pm 0.54$ &
  $1.09\pm 0.066$ & $2.52\pm 0.123$ & 11.1\\
Mrk 169 & $8\pm 2$ & $-0.49\pm 0.07$ &
  $3.17\pm 0.29$ & $4.79\pm 0.29$ & 17.4\\
Haro 28 & $1.5\pm 0.7$ & $-1.11\pm 0.60$ &
  $1.09\pm 0.08$ & $2.29\pm 0.23$ & 10.7\\
Mrk 49 & $4.5\pm 1.5$ & $-0.46\pm 0.52$ &
  $0.724\pm 0.065$ & $0.823\pm 0.140$ & 18.2\\
Haro 9 & $17\pm 5$ & $+0.33\pm 0.41$ &
  $2.63\pm 0.21$ & $4.47\pm 0.27$ & 13.9\\
Mrk 59 & $6.1\pm 1.3$ & $-0.43\pm 0.09$ &
  $1.97\pm 0.12$ & $2.46\pm 0.20$ & 11.6\\
II Zw 70 & $4.4\pm 1.0$ & $-0.26\pm 0.06$ &
  $0.714\pm 0.050$ & $1.24\pm 0.12$ & 17.1\\
Mrk 297 & $22\pm 5$ & $-0.81\pm 0.09$ &
  $6.15\pm 0.31$ & $10.2\pm 0.5$ & 63.0\\
Mrk 313 & $7.5\pm 0.5$ & $-0.63\pm 0.13$ &
  $3.80\pm 0.30$ & $7.40\pm 0.59$ & 30.8\\
Mrk 314 & $3.9\pm 0.7\,^\mathrm{c}$ & --- &
  $1.25\pm 0.10$ & $1.49\pm 0.37$ & 31.1\\
III Zw 102 & $16.7\pm 1.0$ & $-0.62\pm 0.01$ &
  $9.33\pm 0.56$ & $17.8\pm 1.1$ & 25.0\\
\hline
\end{tabular}

\medskip

$^\mathrm{a}$ II Zw 40 and Haro 2 (= Mrk 33) are also
included in \citet{hunt05}'s sample (Table \ref{tab:hunt}).\\
$^\mathrm{b}$ Since $\langle\alpha\rangle$ is not available, we
assume the mean value obtained in \citet{klein91}.\\
$^\mathrm{c}$ Since the flux given by \citet{klein91} is
probably overestimated because of confusion with nearby
sources, we replaced the value with an interferometric flux at
2 cm (15 GHz) given by \citet{deeg97}.\\
Note: Mrk 527 was originally included in the sample of
\citet{klein91}. Because of the suspected contamination of
nearby sources, this galaxy is not included in our sample.
\end{minipage}
\end{table*}

\bsp

\label{lastpage}


\begin{thebibliography}{}
\bibitem[\protect\citeauthoryear{Arimoto, Sofue, \& Tsujimoto}{1996}]{arimoto96}
    Arimoto, N., Sofue, Y., \& Tsujimoto, T. 1996, PASJ, 48, 275
\bibitem[\protect\citeauthoryear{Asano et al.}{2012}]{asano12}
    Asano, R., Takeuchi, T. T., Hirashita, H., \& Inoue, A. K.
    2012, Earth, Planets Space, in press
\bibitem[\protect\citeauthoryear{Baas, Israel, \& Koornneef}{1994}]{baas94}
    Baas, F., Israel, F. P., \& Koornneef, J. 1994, A\&A, 284, 403
\bibitem[\protect\citeauthoryear{Beck, Turner, \& Gorjian}{2001}]{beck01}
    Beck, S. C., Turner, J. L., \& Gorjian, V. 2001, AJ, 122, 1365
\bibitem[\protect\citeauthoryear{Beck et al.}{2002}]{beck02}
    Beck, S. C., Turner, J. L.,
    Langland-Shula, L. E., Meier, D. S., Crosthwaite, L. P., \&
    Gorjian, V. 2002, AJ, 124, 2516
\bibitem[\protect\citeauthoryear{Caldwell \& Phillips}{1989}]{caldwell89}
    Caldwell, N., \& Phillips, M. M. 1989, ApJ, 338, 789
\bibitem[\protect\citeauthoryear{Cannon et al.}{2005}]{cannon05}
    Cannon, J. M., et al.\ 2005, ApJ, 630, L37
\bibitem[\protect\citeauthoryear{Cannon et al.}{2006}]{cannon06}
    Cannon, J. M., et al.\ 2006, ApJ, 647, 293
\bibitem[\protect\citeauthoryear{Chandar et al.}{2005}]{chandar05}
    Chandar, R., Leitherer, C., Tremonti, C. A., Calzetti, D.,
    Aloisi, A., Meurer, G. R., \& de Mello, D. 2005, ApJ, 628, 210
\bibitem[\protect\citeauthoryear{Condon}{1992}]{condon92}
    Condon, J. J. 1992, ARA\&A, 30, 575
\bibitem[\protect\citeauthoryear{Deeg, Duric, \& Brinks}{1997}]{deeg97}
    Deeg, H.-J., Duric, N., \& Brinks, E. 1997, A\&A, 323, 323
\bibitem[\protect\citeauthoryear{de Jong et al.}{1985}]{dejong85}
    de Jong, T., Klein, U., Wielebinski, R., \& Wunderlich, E.
    1985, A\&A, 147, L6
\bibitem[\protect\citeauthoryear{Draine \& Anderson}{1985}]{draine85}
    Draine, B. T., \& Anderson, N. 1985, ApJ, 292, 494
\bibitem[\protect\citeauthoryear{Draine \& Lee}{1984}]{draine84}
    Draine, B. T., \& Lee, H. M. 1984, ApJ, 285, 89
\bibitem[\protect\citeauthoryear{Dumas et al.}{2011}]{dumas11}
    Dumas, G. Schinnerer, E.,
    Tabatabaei, F. S., Beck, R., Velusamy, T., \& Murphy, E. 2011, AJ,
    141, 41
\bibitem[\protect\citeauthoryear{Fazio et al.}{2004}]{fazio04}
    Fazio, G. G., et al.\ 2004, ApJS, 154, 10
\bibitem[\protect\citeauthoryear{Galametz et al.}{2011}]{galametz11}
    Galametz, M., Madden, S. C., Galliano, F., Hony, S., Bendo, G. J.,
    \& Sauvage, M. 2011, A\&A, 532, A56
\bibitem[\protect\citeauthoryear{Galliano et al.}{2005}]{galliano05}
    Galliano, F., Madden, S. C.,
    Jones, A. P., Wilson, C. D., \& Bernard, J.-P. 2005, A\&A, 434, 867
\bibitem[\protect\citeauthoryear{Gao \& Solomon}{2004}]{gao04}
    Gao, Y., \& Solomon, P. M. 2004, ApJ, 606, 271
\bibitem[\protect\citeauthoryear{Gnedin \& Kravtsov}{2011}]{gnedin11}
    Gnedin, N. Y., \& Kravtsov, A. V. 2011, ApJ, 728, 88
\bibitem[\protect\citeauthoryear{Gorjian}{1996}]{gorjian96}
    Gorjian, V. 1996, AJ, 112, 1886
\bibitem[\protect\citeauthoryear{Grassi et al.}{2011}]{grassi11}
    Grassi, T., Krstic, P., Merlin, E.,
    Buonomo, U., Piovan, L., \& Chiosi, C. 2011, A\&A, 533, A123
\bibitem[\protect\citeauthoryear{Helou, Soifer, \& Rowan-Robinson}{1985}]{helou85}
    Helou, G., Soifer, B. T., \& Rowan-Robinson, M. 1985, ApJ, 298,
    L7
\bibitem[\protect\citeauthoryear{Hirashita}{2011}]{hirashita11}
    Hirashita, H. 2011, MNRAS, 418, 828 (H11)
\bibitem[\protect\citeauthoryear{Hirashita \& Ferrara}{2002}]{hirashita02}
    Hirashita, H., \& Ferrara, A. 2002, MNRAS, 337, 921
\bibitem[\protect\citeauthoryear{Hirashita \& Hunt}{2006}]{hirashita06}
    Hirashita, H., \& Hunt, L. K. 2006, A\&A, 460, 67
\bibitem[\protect\citeauthoryear{Hirashita \& Hunt}{2008}]{hirashita_hunt08}
    Hirashita, H., \&
    Hunt, L. K. 2008, Mapping the Galaxy and Nearby Galaxies
    (Ap\&SS Proceedings Ser.), ed.\ K. Wada \& F. Combes, p.\ 333
\bibitem[\protect\citeauthoryear{Hirashita \& Kuo}{2011}]{hirashita_kuo}
    Hirashita, H., \& Kuo, T.-M. 2011, MNRAS, 416, 1340
\bibitem[\protect\citeauthoryear{Hirashita, Tajiri, \& Kamaya}{2002}]{hirashita_etal02}
    Hirashita, H., Tajiri, Y. Y., \& Kamaya, H. 2002, A\&A, 388, 439
\bibitem[\protect\citeauthoryear{Ho, Moran, \& Lo}{2004}]{ho04}
    Ho, P. T. P., Moran, J. M., \& Lo, K. Y. 2004, ApJ, 616, L1
\bibitem[\protect\citeauthoryear{Hopkins, Schulte-Ladbeck, \& Drozdovsky}{2002}]{hopkins02}
    Hopkins, A. M., Schulte-Ladbeck, R. E., \& Drozdovsky, I. O.
    2002, AJ, 124, 862
\bibitem[\protect\citeauthoryear{Hunt, Bianchi, \& Maiolino}{Hunt et al.}{2005}]{hunt05}
    Hunt, L. K., Bianchi, S., \& Maiolino, R. 2005, A\&A, 434, 849
\bibitem[Inoue(2011)]{inoue11} Inoue, A. K. 2011,
    Earth, Planets Space, 63, 1027
\bibitem[\protect\citeauthoryear{Inoue, Hirashita, \& Kamaya}{2000}]{inoue00}
    Inoue, A. K., Hirashita, H., \& Kamaya, H. 2000, PASJ, 52, 539
\bibitem[\protect\citeauthoryear{James et al.}{2002}]{james02}
    James, A., Dunne, L., Eales, S., Edmunds, M. G. 2002, MNRAS,
    335, 753
\bibitem[\protect\citeauthoryear{Johnson \& Kobulnicky}{2003}]{johnson03}
    Johnson, K. E., \& Kobulnicky, H. A. 2003, ApJ, 597, 923
\bibitem[\protect\citeauthoryear{Johnson et al.}{2000}]{johnson00}
    Johnson, K. E., Leitherer, C., Vacca, W. D., \& Conti, P. 2000, AJ,
    120, 1273
\bibitem[\protect\citeauthoryear{Kennicutt}{1998}]{kennicutt98}
    Kennicutt, R. C., Jr.\ 1998, ARA\&A, 36, 189
\bibitem[\protect\citeauthoryear{Klein, Weiland, \& Brinks}{Klein et al.}{1991}]{klein91}
    Klein, U., Weiland, H., \& Brinks, E. 1991, A\&A, 246, 323
\bibitem[\protect\citeauthoryear{Kobulnicky et al.}{1995}]{kobulnicky95}
    Kobulnicky, H. A., Dickey, J. M., Sargent, A. I., Hogg, D. E.,
    \& Conti, P. S. 1995, AJ, 110, 116
\bibitem[\protect\citeauthoryear{Kobulnicky \& Johnson}{1999}]{kobulnicky99}
    Kobulnicky, H. A., \& Johnson, K. E. 1999, ApJ, 527, 154
\bibitem[\protect\citeauthoryear{Kobulnicky, Kennicutt, \& Pizagno}{1999}]{kobulnicky_etal99}
    Kobulnicky, H. A., Kennicutt, R. C., Jr., \& Pizagno, J. L.
    1999, ApJ, 514, 544
\bibitem[\protect\citeauthoryear{Krumholz, Leroy, \& McKee}{Krumholz et al}{2011}]{krumholz11}
    Krumholz, M. R., Leroy, A. K., \& McKee, C. F. 2011, ApJ, 731, 25
\bibitem[\protect\citeauthoryear{Kunth \& \"{O}stlin}{2000}]{kunth00}
    Kunth, D., \& \"{O}stlin, G. 2000, A\&AR, 10, 1
\bibitem[\protect\citeauthoryear{Lada, Lombardi, \& Alves}{2010}]{lada10}
    Lada, C. J., Lombardi, M., \& Alves, J. F. 2010, ApJ, 724, 687
\bibitem[\protect\citeauthoryear{Lasker et al.}{1990}]{lasker90}
    Lasker, B. M., Sturch, C. R., McLean, B. J., Russell, J. L., Jenkner, H., \&
    Shara, M. M. 1990, AJ, 99, 2019
\bibitem[\protect\citeauthoryear{Leitch-Devlin \& Williams}{1985}]{leitch85}
    Leitch-Devlin, M. A., \& Williams, D. A. 1985, MNRAS, 213, 295
\bibitem[\protect\citeauthoryear{Leroy et al.}{2011}]{leroy11}
    Leroy, A. K., et al.\ 2011, ApJ, 737, 12
\bibitem[\protect\citeauthoryear{Lisenfeld et al.}{2007}]{lisenfeld07}
    Lisenfeld, U., et al.\ 2007, A\&A, 462, 507
\bibitem[\protect\citeauthoryear{Lisenfeld \& Ferrara}{1998}]{lisenfeld98}
    Lisenfeld, U., \& Ferrara, A. 1998, ApJ, 496, 145
\bibitem[\protect\citeauthoryear{Lodders}{2003}]{lodders03}
    Lodders, K. 2003, ApJ, 591, 1220
\bibitem[\protect\citeauthoryear{L\'{o}pez-S\'{a}nchez \& Esteban}{2010}]{lopez10}
    L\'{o}pez-S\'{a}nchez, \'{A}. R. \& Esteban, C. 2010, A\&A, 517, A85
\bibitem[\protect\citeauthoryear{MacLaren, Richardson, \& Wolfendale}{1988}]{maclaren88}
    MacLaren, I., Richardson, K. M., \& Wolfendale, A. W. 1988,
    ApJ, 333, 821
\bibitem[\protect\citeauthoryear{Madden}{2000}]{madden00}
    Madden, S. 2000, NewAR, 44, 249 
\bibitem[\protect\citeauthoryear{Mattsson}{2011}]{mattsson11}
    Mattsson, L. 2011, MNRAS, 414, 781
\bibitem[\protect\citeauthoryear{Mauersberger et al.}{1999}]{mauersberger99}
    Mauersberger, R., Henkel, C., Walsh, W., Schulz, A. 1999,
    A\&A, 341, 256
\bibitem[\protect\citeauthoryear{Meier, Turner, \& Beck}{Meier et al.}{2002}]{meier02}
    Meier, D. S., Turner, J. L., \& Beck, S. C. 2002, AJ, 124, 877
\bibitem[\protect\citeauthoryear{Meier et al.}{2001}]{meier01}
    Meier, D. S., Turner, J. L., Crosthwaite, L. P., \& Beck, S. C.
    2001, AJ, 121, 740
\bibitem[\protect\citeauthoryear{Micha{\l}owski et al.}{2010}]{michalowski10}
    Micha{\l}owski, M. J., Murphy, E. J., Hjorth, J., Watson, D., Gall, C., \&
    Dunlop, J. S. 2010, A\&A, 522, A15
\bibitem[\protect\citeauthoryear{Moshir et al.}{1990}]{moshir90}
    Moshir, M., et al.\ 1990, \textit{IRAS} Faint Source Catalog,
    Version 2.0 (Pasadena: Infrared Processing and Analysis Centre)
\bibitem[\protect\citeauthoryear{Nagata et al.}{2002}]{nagata02}
    Nagata, H., Shibai, H.,
    Takeuchi, T. T., \& Onaka, T. 2002, PASJ, 54, 695
\bibitem[\protect\citeauthoryear{Onaka et al.}{2007}]{onaka07}
    Onaka, T., et al.\ 2007, PASJ, 59, S401
\bibitem[\protect\citeauthoryear{Osterbrock}{1989}]{osterbrock89}
    Osterbrock, D. E. 1989,
    Astrophysics of Gaseous Nebulae and Active Galactic Nuclei
    (Mill Valley: University Science Books)
\bibitem[\protect\citeauthoryear{Raiter, Schaerer, \& Fosbury}{2010}]{raiter10}
    Raiter, A., Schaerer, D., \& Fosbury, R. A. E. 2009, A\&A, 523,
    A64
\bibitem[\protect\citeauthoryear{Sage et al.}{1992}]{sage92}
    Sage, L. J., Salzer, J. J., Loose, H.-H., \& Henkel, C. 1992,
    A\&A, 265, 19
\bibitem[\protect\citeauthoryear{Sargent \& Searle}{1970}]{sargent70}
    Sargent, W. L. W., \& Searle, L. 1970, ApJ, 162, L155
\bibitem[\protect\citeauthoryear{Schaerer}{2002}]{schaerer02}
    Schaerer, D. 2002, A\&A, 382, 28
\bibitem[\protect\citeauthoryear{Schmidt \& Boller}{1993}]{schmidt93}
    Schmidt, K.-H., \& Boller, T. 1993, Astron.\ Nachr., 314, 361
\bibitem[\protect\citeauthoryear{Schruba et al.}{2012}]{schruba12}
    Schruba, A., et al.\ 2012, AJ, 143, 138
\bibitem[\protect\citeauthoryear{Shetty et al.}{2011}]{shetty11}
    Shetty, R., Glover, S. C., Dullemond, C. P., \&
    Klessen, R. S. 2011, MNRAS, 412, 1686
\bibitem[\protect\citeauthoryear{Spitzer}{1978}]{spitzer78}
    Spitzer, L. 1978, Physical
    Processes in the Interstellar Medium (New York: Wiley)
\bibitem[\protect\citeauthoryear{Strong et al.}{1988}]{strong88}
    Strong, A. W., et al.\ 1988, A\&A, 207, 1
\bibitem[\protect\citeauthoryear{Sun \& Hirashita}{2011}]{sun11}
    Sun, A.-L., \& Hirashita, H. 2011, MNRAS, 411, 1707
\bibitem[\protect\citeauthoryear{Takeuchi et al.}{2005}]{takeuchi_etal05}
    Takeuchi, T. T., Buat, V., Iglesias-P\'{a}ramo, J., Boselli, A., \&
    Burgarella, D. 2005, A\&A, 432, 423
\bibitem[\protect\citeauthoryear{Takeuchi et al.}{2005}]{takeuchi05}
    Takeuchi, T. T.,
    Ishii, T. T., Nozawa, T., Kozasa, T., \& Hirashita, H. 2005,
    MNRAS, 362, 592
\bibitem[\protect\citeauthoryear{Thuan \& Izotov}{2005}]{thuan05}
    Thuan, T. X., \& Izotov, Y. I. 2005, ApJS, 161, 240
\bibitem[\protect\citeauthoryear{Tully}{1988}]{tully88} Tully, R. B.
    1988, Nearby Galaxies Catalog, Cambridge University Press,
    Cambridge
\bibitem[\protect\citeauthoryear{Turner \& Beck}{2004}]{turner04}
    Turner, J. L., \& Beck, S. C. 2004, ApJ, 602, L85
\bibitem[\protect\citeauthoryear{Turner, Ho, \& Beck}{Turner et al.}{1998}]{turner98}
    Turner, J. L., Ho, P. T. P., \& Beck, S. C. 1998, AJ, 116, 1212
\bibitem[\protect\citeauthoryear{Vacca \& Conti}{1992}]{vacca92}
    Vacca, W. D., \& Conti, P. S. 1992, ApJ, 401, 543
\bibitem[\protect\citeauthoryear{Valiante et al.}{2011}]{valiante11}
    Valiante, R., Schneider, R., Salvadori, S., \& Bianchi, S.
    2011, MNRAS, 416, 1916
\bibitem[\protect\citeauthoryear{van Zee, Skillman, \& Salzer}{1998}]{vanzee98}
    van Zee, L., Skillman, E. D., \& Salzer, J. J. 1998, AJ, 116,
    1186
\bibitem[\protect\citeauthoryear{Vanzi et al.}{2009}]{vanzi09}
    Vanzi, L., Combes, F., Rubio, M., \& Kunth, D. 2009, A\&A, 496, 677
\bibitem[\protect\citeauthoryear{Vanzi et al.}{2008}]{vanzi08}
    Vanzi, L., Cresci, G.,
    Telles, E., \& Melnick, J. 2008, A\&A, 486, 393
\bibitem[\protect\citeauthoryear{Wilson, Walker, \& Thornley}{1997}]{wilson97}
    Wilson, C. D., Walker, C. E., \& Thornley, M. D. 1997, ApJ, 483, 210
\bibitem[\protect\citeauthoryear{Wolfire, Hollenbach, \& McKee}{2010}]{wolfire10}
    Wolfire, M. G., Hollenbach, D., \& McKee, C. F. 2010, ApJ, 716, 1191
\bibitem[\protect\citeauthoryear{Wu et al.}{2008}]{wu08}
    Wu, Y., Charmandaris, V.,
    Houck, J. R., Bernard-Salas, J., Lebouteiller, V., Brandl, B. R., \&
    Farrah, D. 2008, ApJ, 676, 970
\bibitem[\protect\citeauthoryear{Yamasawa et al.}{2011}]{yamasawa11}
    Yamasawa, D., Habe, A., Kozasa, T., Nozawa, T., Hirashita, H.,
    Umeda, H., \& Nomoto, K. 2011, ApJ, 735, 44
\end{thebibliography}
\end{document}